\pdfoutput=1 
\documentclass[
  aps,%
  prx,%
 twocolumn,%
 groupedaddress,% 
 superscriptaddress,%
  showpacs,%
 letterpaper,%
 amsfonts,%
 footinbib,%
 floatfix,%
]{revtex4-2}

\usepackage{hyperref}
\hypersetup{
    colorlinks=true,
    citecolor=blue,
    linkcolor=blue , 
    filecolor=cyan,      
    urlcolor=magenta,
    pdfcreator = {\LaTeX\ and \flqq hyperref\frqq},
}

\RequirePackage{graphicx,amsmath,amssymb,bm}
\usepackage{bbm}
\usepackage{float}
\usepackage{framed} 
\usepackage{amsthm}
\usepackage{caption}
\usepackage{subcaption}
\usepackage[normalem]{ulem}
\usepackage{comment}

\graphicspath{ {./Figs/} } 

\newcommand{\bb}{\mathbb}
\newcommand{\tr}{\text{tr}}

\newcommand{\Ca}{{\cal{C}}}
  
\newcommand{\Aa}{{\cal{A}}}

\newcommand{\bbz}{\bb{Z}}

\def\be{\begin{equation}}
\def\ee{\end{equation}} 
\def\bsh{\begin{shaded}}
\def\esh{\end{shaded}} 
\def\bpm{\begin{pmatrix}}
\def\epm{\end{pmatrix}}

\begin{document}
\title{Gapless symmetry-protected topological phases and generalized deconfined critical points from gauging a finite subgroup} 
\author{Lei Su} 
\affiliation{Department of Physics, University of Chicago, Chicago, Illinois 60637, USA
}
\author{Meng Zeng}
\affiliation{Department of Physics, University of California, San Diego, California 92093, USA}

\begin{abstract} 
Gauging a finite subgroup of a global symmetry can map conventional phases and phase transitions to unconventional ones. In this work, we study, as a concrete example, an emergent $\bbz_2$-gauged system with global symmetry $U(1)$, namely, the $\bbz_2$-gauged Bose-Hubbard model both in 1-D and in 2-D. In certain limits, there is an emergent mixed 't Hooft anomaly between the quotient $\tilde{U}(1)$ symmetry and the dual $\hat{\bbz}_2$ symmetry.  In 1-D, the superfluid phase is mapped to an intrinsically gapless symmetry-protected topological (SPT) phase, as supported by density-matrix renormalization group (DMRG) calculations. In 2-D,   the original superfluid-insulator transition becomes a generalized deconfined quantum critical point (DQCP) between a gapless SPT phase, where an SPT order coexists with Goldstone modes, and a $\tilde{U}(1)$-symmetry-enriched topological (SET) phase. We also discuss the stability of these phases and the critical points to small perturbations and their potential experimental realizations. Our work demonstrates that partial gauging is a simple and yet powerful approach in constructing novel phases and quantum criticalities. 
\end{abstract} 
 
\maketitle 

%%%%%%%%%%%%%%%%%%%%%%%%%%%%%%%%%%%%%%%%%%%

%%%%%%%%%%%%%%%%%%%%%%%%%%%%%%%%%%%%%%%%%%%

\section{Introduction} 
The popular Landau paradigm has been tremendously successful in describing different phases and phase transitions among them. However, more novel phases and phase transitions beyond the traditional paradigm have been found over the past few decades. For example, the deconfined quantum critical point (DQCP) \cite{senthil2023deconfined} between two phases that break different \textit{ordinary } (0-form) symmetries cannot be explained simply by spontaneous symmetry breaking (SSB) from Landau order parameters. Topologically ordered phases \cite{wen2004quantum}, as another example, cannot be captured by SSB of ordinary symmetries. 

It was realized recently that some DQCPs can be explained using mixed 't Hooft anomalies, which can be emergent at low energy  between the two associated symmetries \cite{levin2004deconfined, Metlitski2018}. The concept of 't Hooft anomalies, widely studied in high energy physics, has also found deep and broad applications in condensed matter physics since the discovery of topological insulators or, more generally, symmetry-protected topological (SPT) phases \cite{chen2013,  wen2013, vishwanath2013physics,    kapustin2014anomalous, kapustin2014symmetry, kapustin2014anomalies,  senthil2015symmetry,   witten2016fermion}.  These anomalies characterize global symmetries that cannot be gauged consistently. 
 Related to this work, more recently emergent anomalies have been used to construct gapless SPT phases \cite{keselman2015gapless, scaffidi2017gapless, Verresen2018topology, verresen2021gapless, verresen2021quotient} which are ``intrinsic" in the sense that not only are the topological edge modes robust against the gapless bulk of the system, but also the SPT nature relies crucially on the gaplessness \cite{Thorngren2021,li2022decorated,  Wen2023bulk, li2023intrinsically}. 't Hooft anomalies  thus play an important role in extending the Landau paradigm.

Another perspective in extending the Landau paradigm comes from recent development in expanding the definition of ``symmetries" to generalized symmetries \cite{gaiotto2015generalized, mcgreevy2023generalized, bhardwaj2023lectures} (noninvertible symmetries included \cite{schafer2023ictp, shao2023}) after it was realized that symmetry generators are essentially topological defects. In particular, ordinary (0-form) symmetries, whose charged objects are 0-dimensional, have been generalized to $p$-form symmetries, whose charged objects are $p$-dimensional. Topologically ordered phases can be interpreted as SSB of some higher-form symmetries. Moreover, it was realized that the Higgs phase can be viewed as an SPT phase protected by higher-form symmetries and is stable to weak explicit breaking of these higher-form symmetries \cite{verresen2022higgs, thorngren2023higgs}. 

One more perspective comes from gauging, i.e., coupling systems to dynamical gauge fields. Gauging a theory of matter fields can yield a rich phase diagram. A prominent example is the Fradkin and Shenker model whose phase diagram  can contain a confined phase, a Higgs phase, and a deconfined phase  \cite{Fradkin1979}. The gauging technique can also be used to extract information in the original  system. For example, gauging different SPT phases can lead to distinct topologically ordered phases where quasiparticles have different braiding statistics \cite{levin2012braiding}. 

Anomalies, higher-form symmetries, and gauging form a powerful toolkit and have led to many interesting new discoveries. It is known that coupling a system to a flat gauge field produces a dual higher-form symmetry and that \textit{partially} gauging a discrete symmetry can produce a mixed anomaly between the quotient symmetry and the new dual symmetry \cite{tachikawa2020, wang2018}. It was emphasized in Ref. \cite{su2023boundary} that gauging a finite subgroup is a general approach to construct exotic critical points from ordinary continuous ones. In 1-D, the critical point where the global symmetry is spontaneously broken is mapped to a DQCP between two SSB phases associated with the quotient symmetry and the dual symmetry \cite{zhang2023}. In higher dimensions, it is a generalized DQCP between an ordinary SSB phase and a symmetry-enriched topological (SET) phase. We will analyze the generalized DQCP after partial gauging using the new perspectives from higher-form symmetries and mixed anomalies.

In this work, we study the emergent $\bbz_2$-gauging of a system with global $U(1)$ symmetry. In next section, we describe the general ideas. Starting from Sec.~\ref{sec1d},  we will focus on a concrete model, i.e., the Bose-Hubbard model both in 1-D and in 2-D, coupled to Ising spins on the bonds, where the bosonic parity is effectively gauged.  We adapt the argument in the recently proposed ``Higgs = SPT" paradigm \cite{verresen2022higgs, thorngren2023higgs} to argue that the gauged 1-D superfluid phase is actually an intrinsically gapless SPT phase by considering both the periodic boundary condition (PBC) and the open boundary condition (OBC). The critical low energy theory is a $\bbz_2$-gauged compact boson conformal field theory (CFT). These statements are corroborated by density-matrix renormalization group (DMRG) computations. In Sec.~\ref{sec2d}, we will argue that in 2-D, the superfluid is also a type of gapless SPT where the gaplessness comes from the Goldstone modes and thus the generalized DQCP is between a gapless SPT phase and a SET phase. We also discuss the effect of some perturbations that explicitly break the dual symmetry, and comment on potential realizations in experiments. We conclude our discussion in Sec.~\ref{sec:concl} with some future directions. Some details are presented in the appendices.

\section{General ideas}
Gauging a finite Abelian ordinary (0-form) symmetry in $d$-D space induces a dual $(d-1)$-form symmetry generated by the Wilson operators \cite{gaiotto2015generalized}. The charged objects of the dual $(d-1)$-form symmetry are $(d-1)$-dimensional. One can gauge a finite Abelian normal subgroup $\Gamma$ of the global symmetry $G$ (discrete or continuous), then the global symmetry becomes $G/\Gamma \times \hat{\Gamma}^{(d-1)}$ where $\hat{\Gamma}^{(d-1)} = \hom(\Gamma, U(1))$, the Pontryagin dual of $\Gamma$,  is the dual $(d-1)$-form symmetry. If $G$ is a nontrivial extension of $G/\Gamma$ by $\Gamma$, i.e., $G \ncong  G/\Gamma \times \Gamma$, then there is a mixed anomaly between the $G/\Gamma$ and $\hat{\Gamma}^{(d-1)}$\cite{tachikawa2020}. As a corollary, there is no trivially gapped (i.e., nondegenerate, gapped,
and symmetric under both symmetries) ground state. 

Starting with a general ordinary second order phase transition of Landau type in $d$-D where the global symmetry $G$ is completely spontaneously broken, 
we can obtain a generalized DQCP by  gauging a finite normal subgroup $\Gamma$ of 
$G$ \cite{su2023boundary}. The two phases separated by the generalized DQCP are associated with the SSB of $G/\Gamma$ and $\hat{\Gamma}^{(d-1)}$, respectively. In particular, the SSB of a higher-form symmetry $\hat{\Gamma}^{(d-1)}$ ($d\ge 2)$ leads to a topologically order phase \cite{gaiotto2015generalized, mcgreevy2023generalized}.  For example, we can gauge the $\Gamma =\bbz_2$ subgroup of a $\bbz_4$ clock model in $2$-D where there is an ordinary second order phase transition across which the unbroken $G=\bbz_4$  is completely broken. The transition point now becomes a generalized DQCP between a SSB phase where the quotient $\tilde{\bbz}_2$ is broken and a SET phase enriched by the quotient $\tilde{\bbz}_2$ (see Ref.~\cite{su2023boundary} and also Appendix~\ref{app_a}). 
Using the argument in Refs.~\cite{verresen2022higgs, thorngren2023higgs}, we claim that the quotient $\tilde{\bbz}_2$ SSB phase in fact has boundary modes as long as the dual 1-form $\hat{\bbz}_2$ (as well as the original $\bbz_4$ symmetry) is preserved. If $G$ is continuous, the SSB of $G/\Gamma$ leads to Goldstone modes, the winding number of which is the charge under the dual $\hat{\bbz}_2$ symmetry.  Thus, in the corresponding phase, the boundary modes coexist with the gapless bulk. 

It is even more interesting if 
there is an intermediate phase sandwiched between phases where the global symmetry is preserved or completely broken, such that, after gauging, the dual symmetry and the quotient symmetry are both preserved. For instance, the intermediate critical phase for the 1-D $q$-state clock model with $q\ge 5$ has an emergent $U(1)$ (see Appendix~\ref{app_a}). This is similar to the superfluid phase in the 1-D $XY$ model with global symmetry $U(1)$.  We argue  that the critical phase is an intrinsically gapless SPT phase in the $\bbz_2$-gauged model, described by a symmetry-enriched CFT \cite{verresen2021gapless}. The response action that dictates the symmetry protected edge modes is similar to that in the gauged Ising model  in the  Ref.~\cite{verresen2022higgs}.  However,  as a result of partial gauging, there is a subtle 't Hooft anomaly matching that governs the gaplessness of the SPT phase. 
This idea is not limited to bosonic systems and can be similarly applicable to fermionic systems.  
In our following discussions, we will focus on the bosonic case with $G=U(1)$.

\section{1-D $\bbz_2$-gauged Bose-Hubbard model} 
\label{sec1d}

\subsection{Model}
\label{subsec_1d}
\begin{figure}[tb]
\centering
\includegraphics[width=0.98\columnwidth]{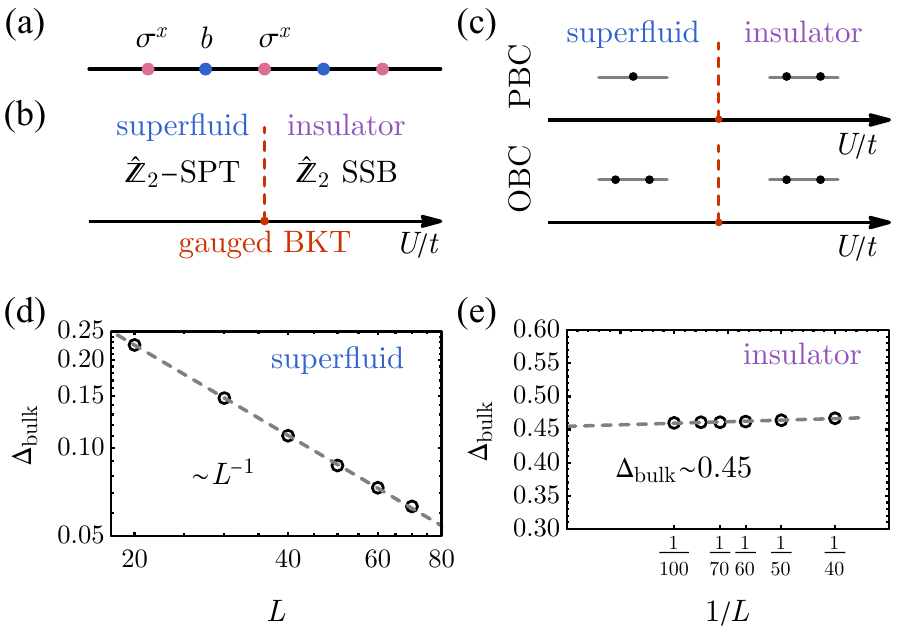} 
\caption{(a) Schematic diagram for the 1-D Bose-Hubbard model (blue sites) coupled to Ising spins (violet bonds).  (b) Schematic phase diagram. The original BKT transition between the superfluid and the insulator phase is enriched to a gauged BKT transition between a gapless SPT superfluid phase, protected by $U(1)$ and $W$, and the insulator phase with $W$ spontaneously broken. (c) Degeneracies in the superfluid and the insulator phase with PBC and OBC, respectively. (d, e) Finite size scaling of the gap $\Delta_{\text{bulk}}$ in both the superfluid ($t =1.0, U =1.0$) and the insulator phase ($t =0.1, U =1.0$). $\Delta_{\text{bulk}}$ is defined to be the gap in spectrum above the (possibly)  degenerate ground states. OBC is used in both cases.} 
\label{fig1d}
\end{figure}

Consider a 1-D Bose-Hubbard model (on the sites, see Fig.~\ref{fig1d}(a)) coupled to Ising spins (on the bonds) as follows:
\begin{align}
     H  & = -t \sum_i b_i^{\dagger} \sigma^z_{i+1/2}  b_{i+1} +U\sum_i n_i (n_i -1)   \nonumber\\
     &  -K\sum_i \sigma^x_{i-1/2} (-1)^{n_i} \sigma^x_{i+1/2},
     \label{eq:ham}
\end{align}  
where $t$ is the hopping, $U >0$ is the on-site Hubbard repulsion, and $n_i=b_i^\dagger b_i$ is the local boson number. In the last term, the Ising spins are coupled to the local boson parity operator $(-1)^{n_i}$. If $K$ is taken to be much larger than the rest of the parameters, then it becomes an emergent parity-gauged Bose-Hubbard model 
\be
     H   = -t \sum_i b_i^{\dagger} \sigma^z_{i+1/2}  b_{i+1} +U\sum_i n_i (n_i -1) 
\ee  
with the gauge constraints
\be 
    G_i=\sigma^x_{i-1/2} (-1)^{n_i} \sigma^x_{i+1/2}=1.
    \label{eq:gauge-constraint1d}
\ee 
Note that in the Hamiltonian, for simplicity, we consider the canonical ensemble where the total boson number $N= \sum_i n_i$ is conserved.  In our following discussion, we consider even system size $L$, regardless of boundary conditions, with one boson per site. This makes the presentation more neat while retaining the essential physics \footnote{Gauging the system with an odd parity is tantamount to gauging with a discrete torsion \cite{vafa1986modular}. Since $H^2(\bbz_2, U(1)) =0$, all gauging processes are equivalent.} .

% ----- F I G U R E -----
% ------------------------------------------------------------------------------

The microscopic model has two symmetries: spin flip  symmetry generated by
\be W=\prod_{i}\sigma_{i+1/2}^z, 
 \label{eq:W-symmetry1d}
\ee 
and boson particle number conservation $U(1)$ symmetry acting as:
\be 
X(\theta)=\prod_{i}e^{i \theta n_i}
\label{eq:xx}
\ee 
with $X(\theta) = X(\theta+2\pi)$. The boson parity 
\be 
P=\prod_i(-1)^{n_i}
\ee
is a subgroup of $U(1)$. 

In the low energy theory, we can interpret the Ising spins as Ising gauge fields. Effectively, bosons on sites are minimally coupled to the Ising gauge field on the bonds.  The boson parity,
viewed as a $\bbz_2$ subgroup of the $U(1)$ symmetry, is gauged, while $W$ can be viewed the dual $\hat{\bbz}_2$ symmetry generated by Wilson loops.  Using PBC, it is easy to see that the UV \textit{physical} symmetry $P$ acts trivially in the IR theory since $P=\prod_i(-1)^{n_i}=\prod_i (\sigma_{i+1/2}^x)^2=1$. This is equivalent to (trivially) projecting out the parity odd sector of the Hilbert space and at the same time adding the twisted sector. Thus, in the IR theory, the original $U(1)$ symmetry effectively reduces to the quotient $\tilde{U}(1)\equiv U(1)/\bbz_2$ symmetry whose action is now 
\be \tilde{X}(\theta)=\prod_i \tilde{X}_i(\theta)\equiv \prod_{i}e^{i \theta n_i/2}.
\label{eq:xtheta}
\ee  Due to the gauge constraints, $\tilde{X}(\theta +2\pi) = \tilde{X}(\theta)$ is satisfied when PBC is used. 
In our following discussion, we will sometimes refer to $\bbz_2$ groups using their generator for simplicity. 

We must distinguish the UV symmetry $W \times U(1)$, where $P$ is a subgroup and hence physical, and the IR symmetry $W \times \tilde{U}(1)$, where the $\bbz_2$ parity is a gauged symmetry. They will play an important role in our later discussion when it comes to the question whether a 't Hooft anomaly is emergent and whether it should be canceled. Also, even though $P =1$ is trivial in the UV because we are considering the case with even $N= L$, it still plays a nontrivial role in the IR. The discussion about the grand canonical ensemble with a finite chemical potential $\mu$ adds more features and is discussed in Appendix~\ref{app_gce}.

\begin{figure}[tb]
\centering
\includegraphics[width=0.98\columnwidth]{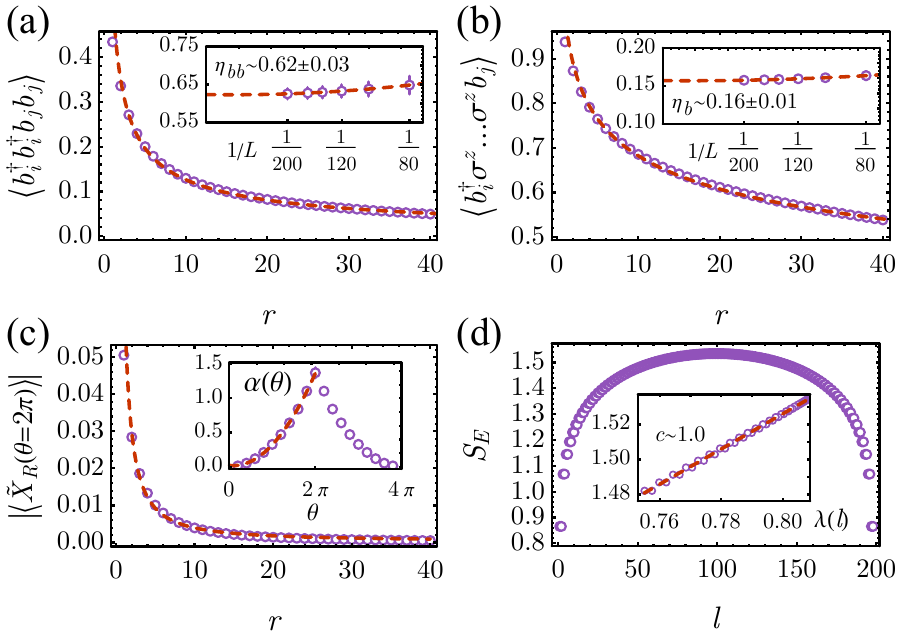} 
\caption{ (a) Boson pair correlation function, which is shown to follow a power law $\langle b_i^\dagger b_i^\dagger b_j b_j\rangle \sim r^{-\eta_{bb}}$ in the superfluid phase. The inset shows the extrapolation of the exponent $\eta_{bb}$ to the thermodynamic limit using finite-size scaling. Error bars are obtained from the upper and the lower bound of the extrapolation. (b) Gauge-invariant boson correlation, which also follows power law $\langle b_i^\dagger \sigma^z...\sigma^z b_j\rangle\sim r^{-\eta_b}$. The inset shows the extrapolation of $\eta_b$. (c) The $\tilde{U}(1)$ disorder operator $|\langle \tilde{X}_R(\theta)\rangle|$ decays as power law $r^{-\alpha(\theta)}$ in $r$. The main plot shows the $\theta=2\pi$ case. The inset shows the $\theta$-dependence of the exponent $\alpha$, which is $4\pi$-periodic. $\alpha$ is symmetric about $2\pi$, and a quadratic fit (dashed line) is performed for the segment from $0$ to $2\pi$.  (d) Subsystem von Neumann entanglement entropy $S_E$ as a function of subsystem size $l$. The inset shows the linear dependence of $S_E$ on $\lambda(l)\equiv \frac{1}{6}\ln \left(\frac{2L}{\pi}\sin\left(\frac{\pi l}{L}\right)\right)$. The central charge $c$, which is given by the slope, is shown to be almost exactly 1. Recently, the entanglement spectrum of gapless SPT phases has also been studied and the entanglement spectrum also has degeneracy \cite{yu2024universal}. All the main plots are for $L=100$, $t=0.5$ and $U=1.0$.} 
\label{fig:boson}
\end{figure}

\begin{figure}[tb]
\centering
\includegraphics[width=0.98\columnwidth]{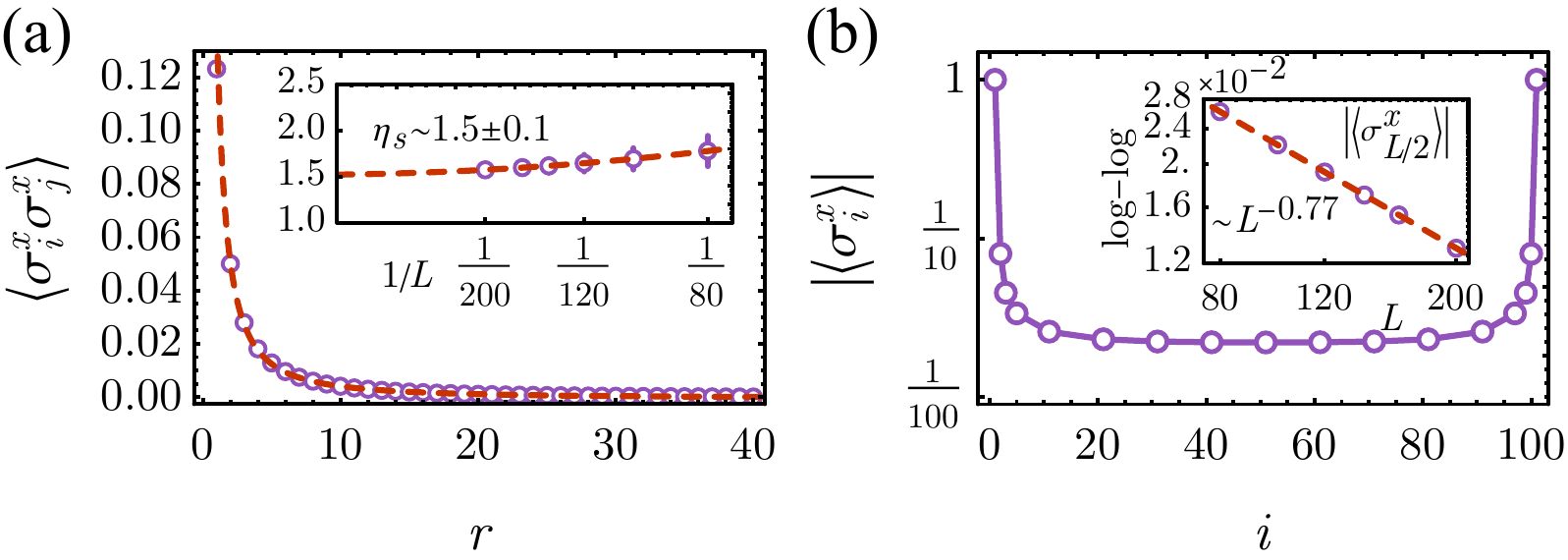} 
\caption{(a) The bulk spin-spin correlation in the superfluid phase with a power-law fit (dash line) $|\langle\sigma_i^x\sigma_j^x\rangle|\sim r^{-\eta_s}$. The inset shows the extrapolation of the critical exponent $\eta_s$ to the thermodynamic limit using finite-size scaling. (b) The magnetization $|\langle\sigma_i^x\rangle|$ in the superfluid phase  across the system with OBC. The spins on the edges are perfectly polarized and the magnetization decays to a small value to the bulk. The inset shows the center spin magnetization $|\langle\sigma_
{L/2}^x\rangle|$ follows a power law decay as system size increases, compatible with the fact that magnetization vanishes when PBC is used. $L=100$, $t=0.5$ and $U=1.0$ for both plots.} 
\label{fig:spin}
\end{figure}

Before coupling to Ising spins, the Bose-Hubbard model can have two phases: a superfluid or a Mott insulator.  The superfluid-insulator transition is a  Berezinskii–Kosterlitz–Thouless (BKT) transition point where the transition is due to fluctuations of vortices in the phase \cite{berezinskii1971destruction, Kosterlitz1973}. The transition occurs around $t/U \approx 0.3$ \cite{Kashurnikov1996Exact, Kuhner1998phases, Kuhner2000}. Note that in the superfluid phase, the $U(1)$ symmetry is not broken due to the celebrated Mermin-Wagner theorem \cite{mermin1966}, but there is a quasi-long range order, where $\langle b^{\dagger}_i b_j \rangle \sim r^{-\eta_b}$ decays algebraically for large $r =|j-i|$ with $\eta_b= \tilde{K}/2$, $\tilde{K}$ being the Luttinger parameter. Similarly, the disorder parameter $|\langle X_R(\theta) \rangle| = |\langle \prod_{i \in R}e^{i \theta n_i}\rangle|$, where $R$ is a line segment with $r =|R| $, decays algebraically. Here, the disorder operator $X_R(\theta)$, unlike the symmetry generator $X(\theta)$ in Eq.~(\ref{eq:xx}), only acts on a subset, $R$, of the total space. If the $U(1)$ symmetry is spontaneously broken, this disorder operator effectively creates two defects at the two ends of $R$. Then the expectation of this operator effectively detects the long-range correlation of the two defects, similar to that in the Ising model \cite{kadanoff1971}. In the 1-D critical superfluid phase, it decays in an algebraic law, similar to the correlation function of two order operators.

Typical phase transitions are insensitive to boundary conditions in the thermodynamical limit.  Gauging the $\bbz_2$ subgroup is equivalent to averaging over untwisted and twisted sectors. Thus, gauging the $\bbz_2$ subgroup does not change the position of critical point in the phase diagram. A continuous phase transition in the gauged model is directly inherited from the ungauged one but with many new features due to the interplay between the quotient $\tilde{U}(1)$ and the dual $\hat{\bbz}_2$ symmetry $W$. The original BKT transition is now gauged [see Fig.~\ref{fig1d}(b)]. 

The intuitions are justified by DMRG computations. 
\begin{comment}
\begin{figure}[tb]
\centering
\includegraphics[width=0.8\columnwidth]{fig-ss-insulator.pdf} 
\caption{$L=100$, $t=0.1$ and $U=1.0$.} 
\label{}
\end{figure}

\begin{figure}[tb]
\centering
\includegraphics[width=0.98\columnwidth]{fig-X-insulator.pdf} 
\caption{$L=100$, $t=0.1$ and $U=1.0$.} 
\label{}
\end{figure}
\end{comment}
After gauging, $\langle b^{\dagger}_i b^{\dagger}_i b_j b_j \rangle$ remains gauge-invariant. The scaling law remains the same as in the ungauged system. $\langle b^{\dagger}_i b_j \rangle$, however, has to be dressed with gauge fields $\sigma^z_i$ to remain gauge-invariant: $\langle b_i^\dagger \sigma_{i+1/2}^z...\sigma_{j-1/2}^z b_j\rangle$. The latter has the same scaling law as $\langle b^{\dagger}_i b_j \rangle$ in the ungauged system.  An example of both order parameters in the superfluid phase is shown in Fig.~\ref{fig:boson}(a-b) where a power-law decay as a function of large $r =|j-i|$ in both parameters can be seen. In the insulator phase, they both decay exponentially to zero. The disorder parameter $|\langle \tilde{X}_R(\theta) \rangle|$ also remains intact. It saturates to a constant in the insulator phase. Its behavior in the superfluid phase is shown in Fig.~\ref{fig:boson}(c) where the angle dependence of $\alpha(\theta)$ is also displayed. $\alpha(\theta)$ has a quadratic dependence on $\theta$. As we will discuss below, this is compatible with  the charge fractionalization in the superfluid phase. Note that even though OBC is used when these quantities are calculated, the bulk behavior is the same as in the PBC case.

On the other hand, the new Ising degrees of freedom also behave differently in the superfluid and the insulator phase. As a result of the emergent gauge constraints, the relation $\langle\sigma_i^x\sigma_j^x\rangle =\langle \prod_{i \in R}(-1)^{n_i} \rangle = \langle \tilde{X}_R(2 \pi) \rangle$ holds in either phase. It relates an ``order parameter" associated with $W$ to a disorder parameter associated with  $\tilde{U}(1)$. As we will discuss soon, this is a manifestation of the emergent mixed anomaly between $\tilde{U}(1)$ and the dual $\hat{\bbz}_2$ symmetry $W$. 
In the insulator case, the magnetization $|\langle\sigma_i^x\rangle|$ in the thermodynamical limit is nonzero, $W$ is spontaneously broken, and the spin-spin correlation  $|\langle\sigma_i^x\sigma_j^x\rangle|$ saturates to a constant. In the superfluid phase, the bulk magnetization vanishes and the bulk spin-spin correlation $|\langle\sigma_i^x\sigma_j^x\rangle|\sim r^{-\eta_s}$ decays algebraically regardless of whether PBC or OBC is used. See Fig.~\ref{fig:spin}(a) for the bulk correlation with OBC \footnote{Note that, due to the edge SSB that we will discuss later, the bulk magnetization remains finite, albeit small, for finite systems. In this plot, we have subtracted the expectation $\langle\sigma_i^x\rangle \langle \sigma_j^x\rangle $ and limited $r$ to be far away from the edges. }. Later on, we will show that these critical exponents are not all independent. In fact, they are compatible with the predictions from a compact boson CFT.  

In our DMRG calculations, the local bosonic Hilbert space dimension is truncated to 5 (beyond which the critical exponents almost saturate).   Bond dimensions less than 400 are sufficient for the results to converge. To obtain the power law exponents, the correlation functions are fitted with a power law decay with $i$ and $j$ far away from both edges. Finite-size scaling is performed with system sizes (number of boson sites) up to $L= 200$.  For each system size, the mean and the error bar are obtained by fitting different segments/bins of data points. The extrapolated mean value is obtained by a quadratic fit against $1/L$. The extrapolated error bar comes from the difference between the extrapolated upper bound and lower bound of the exponent across different system sizes. The DMRG calculations are done using the ITensor package \cite{itensor}.

\subsection{Emergent mixed anomaly}

In the low energy theory, there is an emergent mixed anomaly between  the quotient $\tilde{U}(1)$ and the dual $\hat{\bbz}_2$ symmetry $W$ (considering PBC for simplicity). One manifestation of the mixed anomaly is that $\tilde{U}(1)$ and $W$ cannot be simultaneously realized on-site. Here $\tilde{U}(1)$ and $W$ seem to be realized on-site. However, $\tilde{U}(1)$ is only exact when  the gauge condition in Eq.~(\ref{eq:gauge-constraint2d}) is enforced, i.e., $K\to \infty$, and the Hilbert space then is not a tensor product. We can follow Ref.~\cite{su2023boundary} to eliminate the gauge constraints and find that either $\tilde{U}(1)$ or $W$ is realized non-on-site (see Appendix~\ref{app_eli}).

If we turn on the background gauge field $A^{\tilde{U}(1)}$ and $A^W$ associated with $\tilde{U}(1)$ and $W$ respectively, the mixed anomaly is characterized by a response action in $(2+1)$-D:
$ \omega  =  A^W \cup d A^{\tilde{U}(1)}/4\pi$ 
where $\cup$ is a cup product, a discrete analog of the wedge product of differential forms \footnote{The reader may treat it simply as a wedge product.}. Thus, $\tilde{U}(1)$ and   $W$ cannot be gauged consistently in the $(1+1)$-D system we are studying which may be viewed as the boundary of the $(2+1)$-D bulk (Appendix~\ref{app_ano}). 
The system, viewed as the boundary of the bulk, usually cannot have boundaries because the boundary of a boundary is an empty set. This is not a contradiction because in our work the mixed anomaly is emergent in the low energy sector.  After some suitable modifications of boundary terms, we can put this model on a lattice with open boundary conditions while preserving both symmetries in the Hamiltonian, which plays a crucial role in the ``Higgs = SPT'' argument in Refs.~\cite{verresen2022higgs, thorngren2023higgs} and our argument in the following for the gapless SPT phase.

As a result of the emergent mixed anomaly, the ground state cannot be trivially gapped, meaning that in a gapped phase $W$ must be spontaneously broken since $\tilde{U}(1)$ cannot be spontaneously broken due the Mermin-Wagner theorem. Indeed, $W$ is spontaneously broken in the insulating phase, while in the superfluid phase both $\tilde{U}(1)$ and $W$ are preserved. As we have discussed above, in the superfluid phase, both $\langle b^{\dagger}_i b^{\dagger}_i b_j b_j \rangle$ and  $|\langle \sigma_i^x \sigma_j^x\rangle|$ are shown to have a power law decay  [Fig.~\ref{fig:boson}(a) and \ref{fig:spin}(a)]. These correlated ordering/disordering behaviors are already encoded in the relation $\langle\sigma_i^x\sigma_j^x\rangle =\langle \prod_{i \in R}(-1)^{n_i} \rangle = \langle \tilde{X}_R(2 \pi) \rangle$ we mentioned earlier. This relation implies that $\tilde{U}(1)$ is preserved if  $W$ is spontaneously broken and that $\tilde{U}(1)$ is spontaneously broken if  $W$ is preserved.

Another manifestation of the mixed anomaly is symmetry fractionalization. We take $W$ to act on a segment of sites $R$ instead of the entire chain. This is a disordered operator for the Ising spins. The emergent gauge constraints imply that the string operator has to be dressed with $b$ or $b^{\dagger}$ to act nontrivially on the low-energy sector. In other words, $ 
\langle b_i^\dagger \sigma_{i+1/2}^z...\sigma_{j-1/2}^z b_j\rangle $ is effectively gauge invariant. Note that both $b$ or $b^{\dagger}$ are fractionally charged under $\tilde{X} (\theta)$ in Eq.~(\ref{eq:xtheta}). Similarly, the disorder operator for $\tilde{X} _R(\theta)$ is fractionally charged under $W$. This can be seen from that $\tilde{X}_i (2\pi) = (-1)^{n_i}$ and $ 
\langle \sigma_{i-1/2}^x (-1)^{n_i}... (-1)^{n_j}\sigma_{j+1/2}^x  \rangle
 =1$. Since the edge spins are charged under $W$, linearity implies that the end points of the disorder operator  $ \tilde{X}_R(\theta)  $ are fractionalized.

\subsection{Gapless SPT phase} 
In this section, we show that the superfluid phase is a gapless SPT phase.  

We first adapted the  ``Higgs = SPT" argument in Ref.~\cite{verresen2022higgs} to argue for the existence of edge modes if $W$ and $P$ both commute with the Hamiltonian and an open boundary is chosen such that the emergent gauge constraints in Eq.~(\ref{eq:gauge-constraint1d}) are preserved. Note that we treat $P$ as the physical symmetry in the UV. The emergent gauge constraints force   
$ P=\sigma_{1/2}^x\sigma_{L+1/2}^x$ in the IR. There are different ways to impose boundary conditions on the edges to guarantee the (dynamical) gauge-invariance.  If the edge degrees of freedom are not fixed, Ref.~\cite{verresen2022higgs} argues that $P$ is a physical symmetry, similar to the observation made about asymptotic symmetries in Ref.~\cite{harlow2021symmetries}. In our discussion, we find it more transparent to simply treat $P$ as a UV symmetry. Since the Hamiltonian is local, it must commute with the two $\sigma^x$ individually. The anti-commutativity between $\sigma^x$ with $W$ implies that there are necessarily edges modes if $W$ or $P$ is preserved. To be more explicit, let $|\psi\rangle$ be a ground state of the Hamiltonian that satisfies $W|\psi\rangle= \eta |\psi\rangle$ with $\eta = \pm 1$, then  the state $|\tilde{\psi}\rangle\equiv \sigma^x|\psi\rangle$ is another degenerate state because $W|\tilde{\psi}\rangle=-\eta |\tilde{\psi}\rangle$. If the bulk is non-degenerate, the degeneracy necessarily comes from the edges. In fact, either $W$ or $P$ is spontaneously broken by the edges while the bulk remains gapless. 
These observations can be justified by the DMRG computations. 

First, we compare the degeneracy for PBC and OBC in both the insulator and the superfluid phase. From Fig.~\ref{fig1d}(c), we can see that the ground state in the insulator phase is doubly degenerate, be it with PBC or OBC. This is expected due to the SSB of the dual $\hat{\bbz}_2$ symmetry. There are no edge modes in this phase.  On the other hand, if PBC is used, the ground state of the superfluid phase is unique, while if OBC is used, there is a double degeneracy. This is a result of the SSB of $W$ on the edges we mentioned above. Indeed, we present  the magnetization $|\langle\sigma_i^x\rangle|$ in Fig.~\ref{fig:spin}(b). Even though the bulk magnetization decays to zero in the thermodynamical limit as in the PBC case, the edge spins are clearly polarized. In fact, due to the constraint from $ P=\sigma_{1/2}^x\sigma_{L+1/2}^x = 1$, two edge spins are perfectly correlated.  Note that the degeneracy in this phase is exact even in finite-size systems. This means that the edge modes are strictly localized on the edges, and the edge localization length $\xi_e$, defined as $\mathrm{e}^{-L/\xi_e}\sim\Delta_{\mathrm{bdry}}$, is exactly 0. The wave function can be interpreted as a fixed point SPT state.

Next, we discuss the finite size scaling in the bulk gap $\Delta_{\text{bulk}}$ to show the bulk is indeed gapless in the thermodynamic limit.
As we can see from Fig.~\ref{fig1d}(d), the bulk gap, the first excited state from the doubly degenerate ground state, is inversely proportional to the system size $L$. In the thermodynamic limit, the bulk correlation length $\xi_b$ diverges and the bulk becomes gapless \footnote{Generically, if $\Delta_{\text{bdry}}$ also has a dependence on the system size, the edge degeneracy refers to an exponential decay or at least a decay faster than the bulk gap $\Delta_{\text{bulk}}$ \cite{keselman2015gapless, scaffidi2017gapless, Verresen2018topology, Thorngren2021}.}. This is compatible with the fact that there is a mixed anomaly between $\tilde{U}(1)$ and the dual $\hat{\bbz}_2$ symmetry $W$. As we have already mentioned, both $\tilde{U}(1)$ and  $W$ are preserved, which is supported by algebraically decaying $\langle b_i^\dagger b_i^\dagger b_j b_j\rangle$ [Fig.~\ref{fig:boson}(a)] and $|\langle\sigma_i^x\sigma_j^x\rangle|$ [Fig.~\ref{fig:spin}(a)]. The gap $\Delta_{\text{bulk}}$ in the insulator phase, on the other hand,  remains finite in the thermodynamical limit, as extrapolated by finite-size scaling [Fig.~\ref{fig1d}(e)]. Thus, we have showed that the superfluid phase is a gapless SPT phase.  

We can discuss the effective action of this gapless SPT phase. Let us recall that the microscopic on-site symmetry of the Hamiltonian in Eq.~(\ref{eq:ham}) is $U(1) \times W$. The emergent symmetry acting nontrivially in the IR is $\tilde{U}(1) \times W$. To capture the edge degeneracy,   we may write down a  response action as  \cite{verresen2022higgs}   
\be 
 \alpha = \frac{1}{2} A^W  \cup A^P,
 \label{eq: alpha}
\ee 
where $A^W$ and $A^P$ is the background field of $W$ and $P$ in the spacetime $M$, respectively. If $A^P$ and  $A^W$ are flat, $d A^P =0$ and  $dA^W =0$. Then if $M$ is closed, $\alpha$ is gauge invariant and describes an SPT phase protected by $W$ and $P$. Indeed, if $M$ has a nontrivial boundary, there can be an open Wilson line terminating on the $\partial M$ where  $dA^W\neq 0$. Then the action $S_M = 2\pi \int_M \alpha$  changes by is $\lambda A^W /2$ under the gauge transformation $A^P \to A^P + d\lambda^P$. To compensate this change, there must be edge modes.  

On the other hand, $P$ is a subgroup of $U(1)$. If we turn on a flat background field $A^{U(1)}$, the closedness of $A^{U(1)}$ requires $dA^P = dA^{\tilde{U}(1)}/2\pi  \mod 2 $ (see Appendix~\ref{app_ano}), i.e., $A^P$ may no longer be closed. The action $S_M = 2\pi \int_M \alpha$ now is no longer invariant under $A^W \to A^W + d\lambda^W$ even if $M$ is closed. This is a 't Hooft anomaly between $W$ and $P$ ! 
Since both $W$ and $P$ are UV onsite symmetries, this 't Hooft anomaly must be canceled by some other terms. Luckily, we find that the emergent mixed anomaly can play the role.

Indeed, we have already seen that there is an emergent mixed anomaly between $\tilde{U}(1)$ and $W$. If we denote the $(2+1)$-D bulk as $Y$ such that its boundary is the  $(1+1)$-D spacetime $M$ that we are studying, i.e., $\partial Y  = M$, then the corresponding anomaly action can be written as $  \omega  =  A^W \cup d A^{\tilde{U}(1)}/4\pi$ where $ A^{\tilde{U}(1)}$ and  $A^W$ are extended into $Y$. Note that $\alpha$ and $\omega$ satisfy the anomaly vanishing equation \cite{Thorngren2021}
\be 
\omega  = d \alpha,
\ee 
so the partition function 
\be Z = e^{2\pi i \int_Y \omega  } e ^{-2\pi i \int_M \alpha } \ee 
is anomaly free. In other words, the emergent mixed anomaly compensates the 't Hooft anomaly in $\alpha$. If $M$ has a nontrivial boundary, the gauge invariance argument again justifies the existence of edge modes. Thus, the gapless SPT phase can be captured by $\alpha$ and $\omega$ together.  For more details, see Appendix \ref{app_ano}.  

It is not surprising that Eq.~(\ref{eq: alpha}) is also the effective action of the 1-D SPT phase in Ref.~\cite{verresen2022higgs} where an Ising model is gauged. However, the total symmetry we are considering is $U(1) \times W$ instead of $W \times P$ in that work. Instead of the Higgs phase, our focus here is more on the critical phase. In Ref.~\cite{verresen2022higgs}, they suggested that the critical point is a ``symmetry-enriched quantum critical point" studied in Ref.~\cite{verresen2021gapless} . Here, our critical phase is more closely related to the ``intrinsically gapless SPT phase" proposed in Ref.~\cite{Thorngren2021}. In that work, the (fermionic) parity is gapped by interactions, while in our discussion, the parity is simply gapped by emergent gauging.   The SPT string order parameter in our case is simply $ 
\langle \sigma_{i-1/2}^x (-1)^{n_i}... (-1)^{n_j}\sigma_{j+1/2}^x  \rangle
 =1$ while Higgs order parameter $ 
\langle b_i^\dagger \sigma_{i+1/2}^z...\sigma_{j-1/2}^z b_j\rangle $ vanishes in the thermodynamical limit. Since $H^2( U(1)\times \bbz_2 , U(1))=0$, there is no nontrivial \textit{gapped} SPT phase protected by $U(1)\times W$, based on the complete classification of conventional bosonic SPT phases in 1-D \cite{chen2013}. In other words, if OBC is used for our gapless SPT phase, we cannot gap out the bulk without destroying the edge degeneracy or breaking the total symmetry.

Conceptually, constructing intrinsically gapless SPT phases using partial gauging as we discussed in this work is easier than the approach used in  Ref.~\cite{Thorngren2021}. The mixed anomaly between the quotient symmetry and the dual symmetry is a direct consequence of the nontrivial group extension and does not depend on the details of the Hamiltonian. In Ref.~\cite{Thorngren2021}, the authors considered a fermionic system, but the analysis above is obviously generalizable to fermionic systems even though the fermionic parity cannot be spontaneously broken and spin structures may need to be taken into account. Indeed, a similar analysis can be found in Ref.~\cite{borlar2021gauging}. There, they gauged the fermionic parity of a free spinless fermionic chain. Hubbard interactions can also be added. As long as there is no other SSB order, the gauged Luttinger liquid of spinless fermions with $W$ preserved is a gapless SPT phase. Spinful fermions can also be studied. Note that gauging the fermionic parity produces a bosonic theory that does not depend on spin structures. In particular, gauging the parity of a 1-D Dirac fermion theory yields the $XY$ model with a BKT transition \cite{karch2019}.

Even though we have focused on a canonical ensemble with even parity, the analysis carries over to a grand canonical ensemble. Then both even and odd parity sectors should be taken into account, especially when OBC is used. The essential physics stays almost unchanged. For example,  the ground state degeneracy for both PBC and OBC is the same as in Fig.~\ref{fig1d}(b). The anomaly analysis is similar. For more details, see Appendix~\ref{app_gce}.

\subsection{Conformal field theory}
Since the BKT transition can be described by a compact boson CFT \cite{ginsparg1988applied}, we expect that the gauged BKT transition and the  superfluid phase is also captured by gauging the $\bbz_2$ symmetry of the compact boson CFT which is also a compact boson CFT. Indeed, the compact boson CFT contains two global $U(1)$'s, one associated with momentum and the other with winding. They are dual to each other and there is a mixed anomaly between them.  Gauging a $\bbz_2$ subgroup of a compact boson CFT not only changes the radius of the compactification, but also maps order operators to disorder operators and vice versa \cite{su2023boundary, bhardwaj2023lectures}. The $\bbz_2$-charged sectors and the $\bbz_2$-twisted sectors are exchanged under this operation. The states in the $\bbz_2$-twisted sectors are charged under the dual $\hat{\bbz}_2$ symmetry. This dual $\hat{\bbz}_2$ symmetry can be viewed as a subgroup of the $U(1)$ symmetry associated with winding.

This expectation again can be verified by the DMRG results. We first check the central charge $c$ in the superfluid phase in the gauged system. Indeed, as shown in Fig.~\ref{fig:boson}(d), the subsystem entanglement entropy  $S_E = -\tr(\rho_R \ln \rho_R)$, associated with the reduced density matrix $\rho_R$ of a subsystem $R$,  scales linearly with the factor $\lambda(l)\equiv \frac{1}{6}\ln \left(\frac{2L}{\pi}\sin\left(\frac{\pi l}{L}\right)\right)$ \cite{calabrese2009entanglement}.  Here $L$ is the total system size of the open chain and $l$ is the size of the subsystem $R$ on one side of the entanglement cut. The slope gives us the central charge $c \sim 1$, the nominal central charge of a compact boson CFT.

Next, we identify the microscopic operators with the primary vertex operators.  Before gauging, the (Euclidean) action is given by 
\be 
S = \frac{1}{4\pi} \int dz d\bar{z} \partial_z \phi \partial_{\bar{z}} \phi,
\ee
where $z = \exp(\tau +i x)$, and the free boson field $\phi$ is compactified on a circle of radius $R$, i.e., $\phi(z, \bar{z}) \sim \phi(z, \bar{z}) +2\pi R$. Split $\phi(z, \bar{z})$ into the left-moving and the right-moving components: $\phi(z, \bar{z}) = X_L(z)+ X_R(\bar{z})$. Then the local primary operators are:
\begin{align}
 V_{n, w}&(z, \bar{z} )  \\
 &= \exp\left[ i \left(\frac{n}{R} + w R\right) X_L(z)+i \left(\frac{ n}{R} -   w R\right) X_R(\bar{z}) \right],  \nonumber 
\end{align}
where $n \in  \bbz$ and $w\in \bbz$ are the momentum number and the winding number, respectively.  After gauging, $R \to R/2$, which is equivalent to redefining $n$ and $w$:  $n \in \frac{1}{2} \bbz$ and $w\in 2 \bbz$ and $\phi(z, \bar{z}) \sim \phi(z, \bar{z}) +4\pi R$ while fixing $R$ \cite{ji2020top, su2023boundary}. The conformal weights of $V_{n, w}$ are 
\be 
h_{n, w} = \frac{1}{4}  \left(\frac{n}{R} + w R\right) ^2, \quad \bar{h}_{n, w} = \frac{1}{4}  \left(\frac{n}{R} - w R\right) ^2,
\ee
and conformal dimensions
\be \Delta_{n, w} =  h_{n, w} +\bar{h}_{n, w}= \frac{1}{2}(\frac{n^2}{R^2} + w^2 R^2).
\ee 
At a generic radius, the CFT has global symmetry $U(1)_n \times U(1)_w$ which act on $X_{L/R}$ as :
\begin{align}
U(1)_n: & X_{L/R}(z)  \to X_{L/R}(z) +  R  \theta_n, \nonumber \\
U(1)_w: & X_{L/R}(z)  \to X_{L/R}(z) \pm \frac{1}{4R} \theta_w, 
\end{align}
where  $\theta_{n/w} \sim \theta_{n/w}+2\pi $.
On the gauged vertex operators $V_{n, w}$, they act as 
\begin{align}
U(1)_n: &V_{n, w}  \to  e^{i 2 n\theta_n} V_{n, w},\nonumber \\
U(1)_w: &  V_{n, w}  \to   e^{i w \theta_w/2} V_{n, w}.
\end{align}
In our case, we can identify $\tilde{U}(1)$ with $U(1)_n$ and identify $\hat{\bbz}_2$ with the $\bbz_2$ subgroup of $U(1)_w$. It is straightforward to see that  $b^{\dagger} b^{\dagger}$ (or $b b$) can be identified with $V_{2, 0}$ and $\sigma^x$ with $V_{0, 1/2}$. The corresponding conformal dimensions are thus $2/R^2$ and $R^2/2$ respectively. Meanwhile, the nonlocal operator $b^{\dagger}$ (or $b$) corresponds to the nontrivial local operator with the lowest scaling dimension $V_{1, 0}$ before gauging. Its scaling dimension should be $1/2R^2$.  This is indeed supported by DMRG. 

In the DMRG calculations, we choose a sample point in the superfluid phase: $t =0.5, U =1.0$. As shown in  Fig.~\ref{fig:boson} and Fig.~\ref{fig:spin}, the scaling dimension of $b^{\dagger} b^{\dagger}$ is $\eta_{bb}\sim 0.62$ and the scaling dimension of the scaling dimension of $\sigma^x$ is $\eta_{s}\sim 1.5$.  The nonlocal correlation $\langle b_i^\dagger \sigma^z...\sigma^z b_j\rangle\sim r^{-\eta_b}$ yields a scaling dimension $\eta_{b}\sim 0.16$. They are all compatible with $R^2 \sim 3.1$.  Note also that the conformal dimensions of the order/disorder operators scale quadratically with the charges $n$ and $w$. Previously we have seen that the end points of the disorder operator $ \tilde{X}_R(\theta)  $
are fractionalized charged under the dual $\hat{\bbz}_2$. Assuming linearity in charge fusion, we may \textit{formally} assign a charge proportional to $\theta$  to the end points when $0\le \theta\le 2\pi$. Thus we may expect that the conformal dimension of $ \tilde{X}_R(\theta)  $ should be proportional to $\theta^2$ in the interval. This is compatible with the quadratic fit in the inset of Fig.~\ref{fig:boson}(c).

\subsection{Perturbations}
In the discussions above, 
we have argued that the superfluid is a gapless SPT is protected by $U(1)$ and $W$.  As long as the Hamiltonian (and the boundary conditions) preserve $U(1)$  and the dual $\hat{\bbz}_2$ symmetry $W$, the edge degeneracy is protected. For instance, as verified by DMRG, adding a term  $\sum \sigma^z_{i-1/2} \sigma^z_{i+1/2}$ does not lift the degeneracy.  Adding a small perturbation $h_x \sum \sigma^x_{i+1/2}$ however breaks this symmetry. The edges open up a small gap that scales as $1/L$ and are no longer degenerate (see Appendix \ref{app-perturb} for more details). The situation is different in higher dimensions when the protecting symmetries include higher-form symmetries. Breaking higher-form symmetries explicitly may not lift the edge degeneracy. Furthermore, we could introduce another type of perturbation $h_z\sum\sigma^z_{i+1/2}$, which preserves the $W$ symmetry, but violates the effective gauge constraint from the large $K$-term. When $h_z\ll K$, the gauge constraint will still be effectively enforced at low energy, and exact edge degeneracy remains. Only when $h_z<K$ becomes sufficient large, does the degeneracy become lifted, and the edge modes become exponentially localized instead of exactly localized (See Appendix \ref{app-perturb} for more details).

\section{2-D $\bbz_2$-gauged Bose-Hubbard model} 
\label{sec2d}
\begin{figure}[tb]
\centering
\includegraphics[width=0.92\columnwidth]{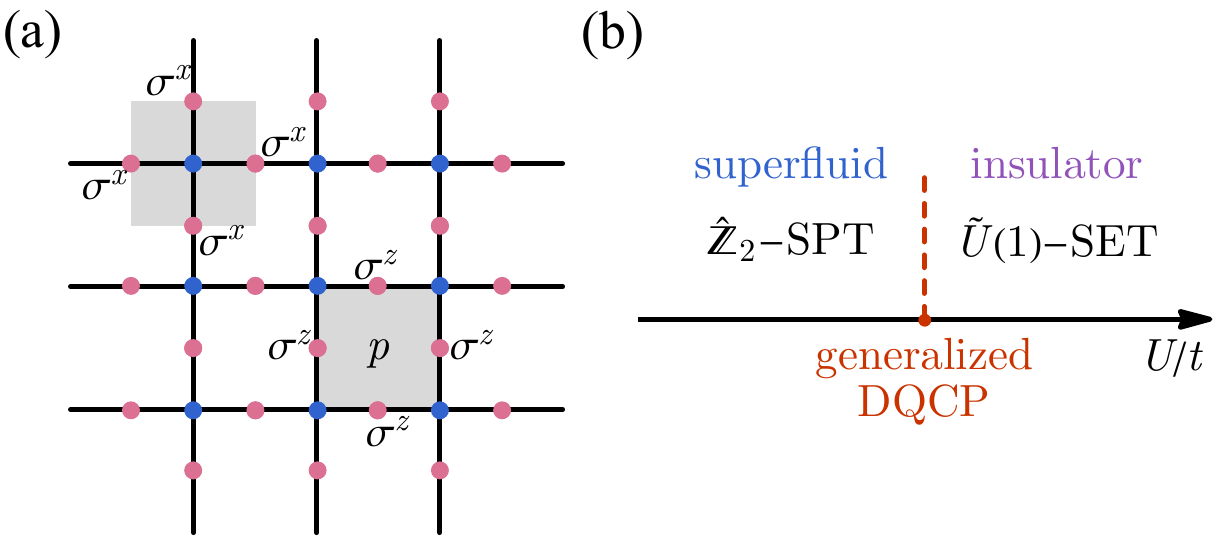} 
\caption{(a) Schematic diagram for the 2-D Bose-Hubbard model (blue sites) coupled to Ising spins (violet bonds). A star operator related to the gauge constraints in Eq.~(\ref{eq:gauge-constraint2d}) and a plaquette operator are highlighted. (b) Schematic phase diagram. The gapless SPT in the superfluid phase where $\tilde{U}(1)$ is spontaneously broken and the SET enriched by $\tilde{U}(1)$ in the insulator phase are separated by a generalized DQCP.}

\label{fig2d}
\end{figure}

Having considered the 1-D case, we can generalize the analysis to higher dimensions.  In this section, we consider the emergent $\bbz_2$-gauged Bose-Hubbard model on a 2-D square lattice [Fig.~\ref{fig2d}(a)]:

\begin{align}
    H &= -t \sum_{  i, j \in \partial e_{ij} } b_i^{\dagger} \sigma^z_{e_{ij}} b_j +U\sum_i n_i (n_i -1) - \mu \sum_i n_i,\nonumber \\
     &  - \frac{1}{g}\sum_p \prod_{e \in \partial p} \sigma_e^z - g  \sum_e \sigma_e^x,
     \label{eq2d}
\end{align}
with the gauge constraints
\be 
    G_i=\prod_{i \in \partial e} \sigma_e^x (-1)^{n_i}=1.
    \label{eq:gauge-constraint2d}
\ee 
Here $e_{ij}$ represents the bond connecting site $i$ and site $j$, and $p$ represents any plaquette of the lattice. 
Note that in order to capture the Higgs phase, we turn on the chemical potential $\mu$ and consider the grand canonical ensemble. The 1-D case is briefly discussed in Appendix~\ref{app_gce}. 
Similar to the 1-D version, we view the gauge constraints to be energetically enforced. 

We first let $g \to 0$ so that the zero-flux (flatness) condition  $\prod_{e \in \partial p}\sigma_e^z =1 $ is enforced and the transverse field term is dropped, giving rise to an emergent 1-form symmetry
\be
    W=\prod_{e\in \gamma}\sigma_e^z,
    \label{eq:W-symmetry}
\ee
where $\gamma$ is a loop running along the bonds of the lattice. Nonzero $g$ perturbations will be discussed later.  There is also a $\tilde{U}(1)$ with
 $  \tilde{X}(\theta) =\prod_{i}e^{i \theta n_i/2}$ satisfying $\tilde{X}(\theta +2\pi) = \tilde{X}(\theta)$. This model has been studied before in, e.g., Ref.~\cite{Swingle2012} from a different perspective. In their studies, the boson field $b$ is not fundamental but emergent as a result of fractionalization. In our following discussion, we will emphasize more on higher-form symmetries and anomalies. As it is hard to study large systems using DMRG, we will focus on the theoretical analysis, although some results have been checked already in small systems using DMRG.

It is well-known that
there is a second order superfluid-insulator transition in the pure Bose-Hubbard model (before coupling the Ising model) by tuning the ratio $U/t$. Unlike the 1-D version, the global $U(1)$ symmetry is spontaneously broken in the superfluid phase.  For simplicity, we may also assume that the chemical potential $\mu$ has been tuned such that the boson filling is an integer.  In the gauged model, the zero-flux condition ensures the flatness of the gauge field,  killing all local dynamics but the topological degrees of freedom in $\sigma^z$. Thus a continuous  phase transition in the gauged model is directly inherited from the ungauged one but with many new features due to the interplay between the quotient $\tilde{U}(1)$ and the dual $\hat{\bbz}_2$ 1-form symmetry $W$.

% ------------------------------------------------------------------------------
% ----- F I G U R E -----
% ------------------------------------------------------------------------------
% ------------------------------------------------------------------------------

\subsection{Emergent mixed anomaly}
Since $U(1)$ is a nontrivial extension of $\tilde{U}(1)$ by $\bbz_2$, there is a mixed anomaly between $\tilde{U}(1)$ and the $\hat{\bbz}_2$ 1-form symmetry $W$(see Appendix~\ref{app_ano}).  Let us denote the $2$-D system as $M$ (without boundaries) and view it as a boundary of a $3$-D bulk $Y$. The mixed anomaly is captured by a (3+1)-D SPT bulk protected by the generalized symmetry $\tilde{U}(1) \times W$.     If we turn on the 1-form background gauge field $ A^{\tilde{U}(1)}$ of $\tilde{U}(1)$ and the 2-form background gauge field $A^W$ of $W$ and extend them into $Y$, then the anomaly action is given by 
\be 
S_Y=  \frac{i}{2} \int_Y A^{W } \cup d A^{\tilde{U}(1)} .  
\label{eq:anomaly}
\ee
Here, $ A^{\tilde{U}(1)}$ is compact and periodic in $2\pi$, and $A^{W}$ takes value in $\bbz_2$.
This action is not gauge invariant under the gauge transformations of $A^{W}$ in the presence of boundary $M$, a manifestation of the mixed anomaly between $\tilde{U}(1)$ and $W$ in the boundary theory which implies that it is impossible to gauge both symmetries consistently. In fact, the mixed anomaly reduces to that in the case of gauging the $\bbz_2$ subgroup of $\bbz_4$ \cite{ su2023boundary, moradi2023symmetry}. Analogous to the 1-D case, the end points of the disordered operators of one symmetry are fractionally charged under the other, which can be seen directly from $\langle b_i^\dagger \sigma_{i+1/2}^z...\sigma_{j-1/2}^z b_j\rangle$ and  $|\langle \tilde{X}_R(\theta) \rangle|$.

Similar to the 1-D case, the ground state of the system cannot be trivially gapped (i.e., nondegenerate, gapped, and symmetric under both symmetries). This consequence strongly constrains the phase diagram. As we will argue in this work, the critical point inherited from the ordinary superfluid-insulator transition now becomes a generalized DQCP between a gapless SPT phase where $\tilde{U}(1)$ is spontaneously broken and a SET phase where the dual $\hat{\bbz}_2$ symmetry is spontaneously broken [Fig.~\ref{fig2d}(b)].

\subsection{Gapless SPT phase }  
We now combine the emergent mixed anomaly with the ``Higgs\ =\ SPT'' argument in Ref.~\cite{verresen2022higgs} to argue in two steps that the superfluid phase after gauging becomes a gapless SPT phase.

As the first step, we show that if the 1-form symmetry $W$ is not spontaneously broken, then the ground state is a gapless SPT phase. The gaplessness is a direct consequence of the mixed anomaly: if $\tilde{U}(1)$ is also preserved, then the system is critical; on the other hand, if $\tilde{U}(1)$ is spontaneously broken, there will be Goldstone bosons.  When the system has no boundary, the gauge condition Eq.~(\ref{eq:gauge-constraint2d}) implies that the boson parity symmetry $P$ is trivial in the low energy  effectively gauged system, $P=\prod_i(-1)^{n_i}=\prod_e (\sigma_e^x)^2=1$. However, when there is an open boundary (which preserves necessary symmetries), $P$ becomes non-trivial because the spin operators on the boundary are not canceled, $P=\prod_{e\in \text{bdry}}\sigma_e^x$. Using some rough terms, we may state that in this case $P$ is not ``completely" gauged even in the low-energy sector.  It is manifested in the existence of ``half" string operators $\sigma_{1/2}^z...\sigma_{j-1/2}^z b_j$ with one end terminating on the boundary  which acts nontrivially in the low energy sector. Nevertheless, boson creation/annihilation operators still have to be attached by a string of gauge field $\sigma^z$. In particular, if the  string operator does not end on the boundaries, it has to end on a creation or an annihilation operator to ensure a nontrivial action in the low energy sector.  This means that the emergent mixed anomaly is still playing its due role and the ground state still cannot be trivially gapped. 

\begin{figure}[tb]
\centering
\includegraphics[width=\columnwidth]{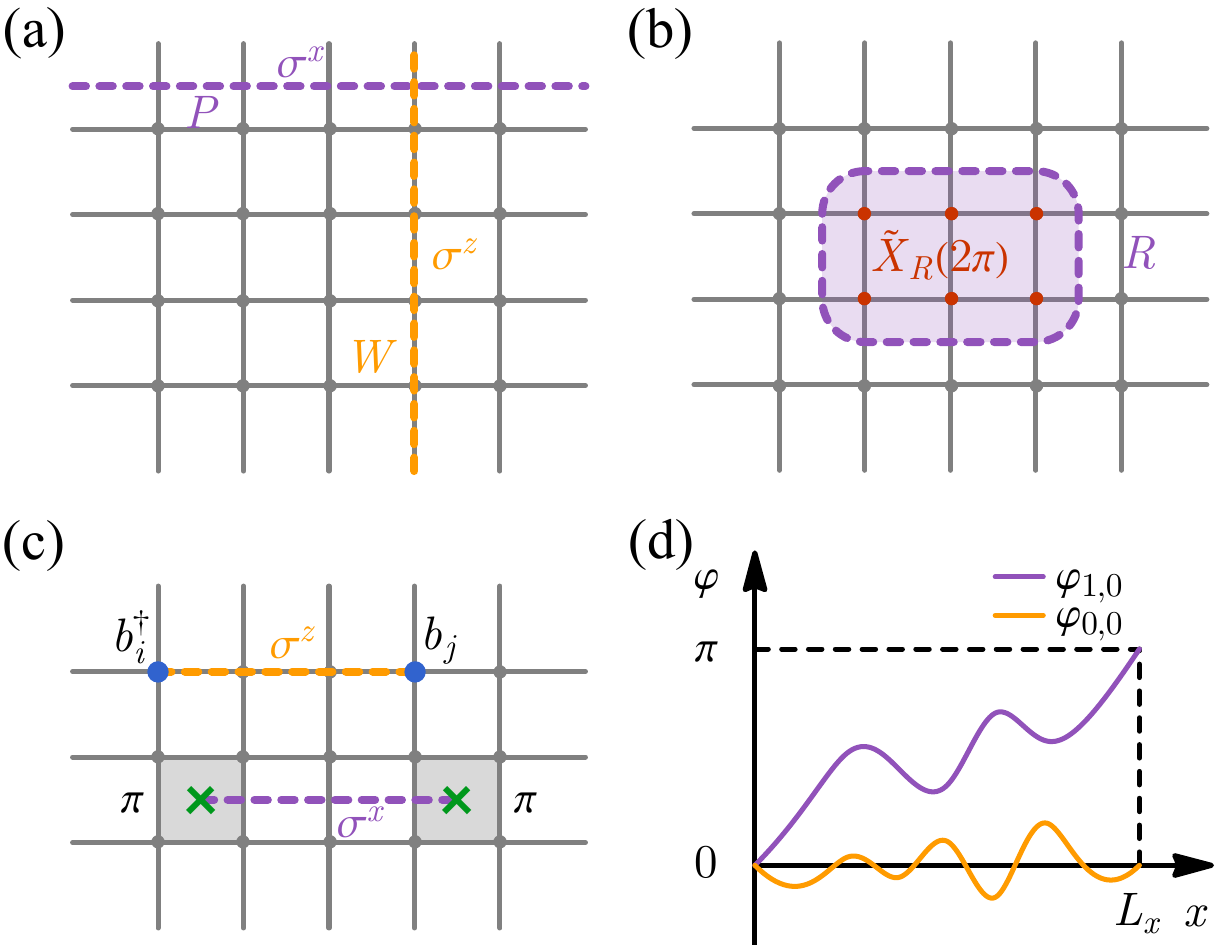} 
\caption{(a) Action of boson parity $P$, effectively a 't Hooft line (purple), and a Wilson line $W$ (orange) terminating on the boundary (top) of a semi-infinite system. The other end of $W$ terminates either in the bulk or at infinity. The anti-commutativity of $P$ and  $W$  forces a SSB on the boundary. (b) Disorder operator $\tilde{X}_R(\theta=2\pi)$ supported on sites (highlighted in red) inside a region $R$, which is the same with with the 't Hooft loop operator, $\prod_{e\in \partial R}\sigma_e^x$, supported on the (dual lattice) boundary of $R$. (c) Gauge invariant Wilson line $W$ attached to boson operators (Higgs order operator), and a 't Hooft line connecting two $\pi$-vortices (which is suppressed by the zero-flux condition). (d) Example of two topologically distinct phase modes with winding number 1, $\varphi_{1,0}$ (red), and winding number 0, $\varphi_{0,0}$ (blue) along the $x$-direction. 0 and $L_x$ are identified.}
\label{fig:PW}
\end{figure}

To show that the state is SPT, we place it on lattice with open boundaries that preserve the dual $\hat{\bbz}_2$ symmetry $W$. We consider a half-plane geometry with an infinitely long boundary [Fig.~\ref{fig:PW}(a)], where the ``rough" boundary has dangling bonds sticking out so that properly modified local gauge constraints Eq.~(\ref{eq:gauge-constraint2d}) are still well-defined \cite{bravyi1998quantum}. Then similar to the 1-D case, we can argue that if $W$ is preserved, there will necessarily be a boundary degeneracy as follows.    We choose the symmetry generator of $W =\prod_{e\in \gamma}\sigma_e^z$  to be a Wilson line with one end terminating on the boundary and the other end either extending to infinity or terminating on a boson creation/annihilation operator. With $P$ and $W$ preserved in the bulk, the anti-commutativity of $P$ and $W$  implies a ground state degeneracy, which in this case necessarily comes from the boundary. This defines an SPT phase. In fact, the Higgs condensate $\langle b_i^\dagger \sigma_{i+1/2}^z...\sigma_{j-1/2}^z b_j\rangle$ can be viewed as an SPT string order parameter here.

% ----- F I G U R E -----
% ------------------------------------------------------------------------------

% ------------------------------------------------------------------------------

As the second step, we show that $W$ is preserved in the superfluid  phase. To this end, we show that a 't Hooft loop operator, charged under $W$, satisfies the non-perimeter law. We take the 't Hooft loop (defined on the dual lattice) to be $\prod_{e\in \partial R}\sigma_e^x$ where $R$ is an arbitrary large connected area with perimeter $l=|\partial R|$, shown in Fig.~\ref{fig:PW}(b) (we ignore the contributions from corners for simplicity). Then the gauge condition Eq.~(\ref{eq:gauge-constraint2d}) implies that 
$  |\langle\prod_{e\in \partial R}\sigma_e^x\rangle| =|\langle \prod_{i\in R}(-1)^{n_i}\rangle|=|\langle   \prod_{i\in R} e^{i\pi n_i}\rangle |=|\langle   \tilde{X}_{R}(2\pi) \rangle|$. Since $\tilde{X}_{R}(2\pi)$ is a disorder operator for $\tilde{U}(1)$, we can see that the order parameter of the dual $\hat{\bbz}_2$ and the disorder parameter of the $\tilde{U}(1)$ are directly related. We interpret this relation as a direct manifestation of the mixed anomaly: if one symmetry is spontaneously broken, then the other is preserved. There is one subtlety here.  In the insulating phase, $|\langle   \tilde{X}_{R}(\theta)  \rangle|$ and $  |\langle\prod_{e\in \partial R}\sigma_e^x\rangle|$ both satisfy the perimeter law $e^{-\beta(\theta) l}$, where $\beta(\theta)$ is independent of $l$. Thus, $\tilde{U}(1)$ is preserved while the dual $\hat{\bbz}_2$ is spontaneously broken in this phase. It is possible to absorb the dependence on $\partial R$ in the perimeter law by adding a local counterterm such that both quantities approach a constant for large smooth $\partial R$ \cite{hastings2005}. This is the SET phase in Fig.~\ref{fig2d}(b) which we will discuss more later.  On the other hand, in the superfluid phase, $|\langle   \tilde{X}_{R}(\theta)  \rangle|$ satisfies the scaling $  \sim e^{-\alpha(\theta) l\ln l}$ where $\alpha (\theta)$ is independent of $l$ \cite{lake2018higher, wang2021scaling}, weaker than the area law. Consequently, the scaling of $  |\langle\prod_{e\in \partial R}\sigma_e^x\rangle|$ is also weaker than the area law. Since it is strictly stronger than the perimeter law, it is impossible to renormalize the scaling law to a constant for large smooth $\partial R$. Thus, we claim that $W$ is unbroken.  

Combining the two steps, we conclude that the superfluid is mapped to a gapless SPT phase protected by $W$ and $P$ where $\tilde{U}(1)$ is spontaneously broken.  The gaplessness comes from the Goldstone bosons.

Similar to the 1-D case, we may write down the term in the effective action that dictates the existence of edge modes \be 
 \alpha  = \frac{1}{2} A^W  \cup  A^P, 
 \label{eq:ano2d}
\ee 
where $A^W$ is the  2-form  background gauge field of $W$ and $ A^P$ is the  1-form  background gauge field  of $P$. The anomaly in $\alpha$ when $A^{\tilde{U}(1)}$ is turned on is again canceled by 
the anomaly action in Eq.~(\ref{eq:anomaly}).
In some rough sense, $\alpha$ dictates the existence of edge modes when $dA^{\tilde{U}(1)}/2\pi = 0$ mod 2, while the mixed anomaly governs the SSB of $\tilde{U}(1)$ in the superfluid phase.

\subsection{Excitations in the superfluid phase}
We have argued that the superfluid phase is a gapless SPT phase.  Here we discuss the excitations in the phase:  vortices,  Goldstone bosons, and domain walls. Since the $\bbz_2$-gauging process amounts to projecting out the $\bbz_2$-charged sector and adding the twisted sector to the theory and the process relates some order parameters to disorder parameters \cite{su2023boundary, moradi2023symmetry}, many physical properties of this new gapless SPT phase can be inferred directly from its superfluid parent state. As we have already seen above, the disorder parameter $|\langle \hat{X}_R(\theta)\rangle|$ for $\tilde{U}(1)$ is a (fractional) order parameter for the dual $\hat{\bbz}_2$. Similarly, $|\langle b_i^{\dagger}(\prod_{e\in\gamma_{ij}}\sigma_e^z) b_j\rangle| $ where $\gamma_{ij}$ connects sites $i$ and $j$ [Fig.~\ref{fig:PW}(c)],  viewed as a disorder parameter for the dual $\hat{\bbz}_2$, serves as a (fractional) order parameter for $\tilde{U}(1)$.

To guarantee the exactness of the 1-form symmetry, we imposed the zero-flux condition  $\prod_{e \in \partial p}\sigma_e^z =1 $, which suppresses all (dynamical) $\pi$-vortices excitations. Equivalently, an open 't Hooft line which would end on a pair of $\pi$-vortex [Fig.~\ref{fig:PW}(c)] are also suppressed. In the Higgs phase,  the condensate phase $\varphi$ is locked to the vortices. The dual 1-form symmetry measures the $\bbz_2$ winding of condensate phase $\varphi$, which takes values in $n\pi$ for integer $n$ modulo 2, along noncontractible cycles, which is equivalent to insertions of $\pi$-fluxes across noncontractible cycles [see Fig.~\ref{fig:PW}(d)]. We should compare the scenario with that in the ``parent" superfluid phase before gauging where there is an emergent 1-form $U(1)$ symmetry in the low energy sector \cite{gaiotto2015generalized, Dela2020super,  pace2023emergent}. Charged objects of the emergent $U(1)$ are the winding of $\varphi$, taking values in $2 n\pi$ for integer $n$. $2\pi$-vortices explicitly breaks this emergent dual $U(1)$ 1-form symmetry. However, since they are neutral under the dual $\hat{\bbz}_2$ 1-form symmetry $W$ after the effective gauging, they do not destroy the exactness of the dual $\hat{\bbz}_2$ 1-form symmetry. 

Being put on a torus, the gapless Goldstone modes can be effectively decomposed into two parts: $\varphi = \varphi_{n, m} + \delta \varphi$, where the first term denotes winding of $n\pi$ and $m\pi$  along the two noncontractible cycles separately and the second term is the small fluctuation with respect to this configuration. Thus the topology of the Goldstone modes can be captured by the 1-form charges. It is very tempting to compare it with the topology of Goldstone modes after a continuous symmetry $G$ is spontaneously broken to a subgroup group $H$ \cite{else2021top}. There, the symmetry protection/enrichment of the Goldstone modes is discussed with respect to the residual symmetry $H$ (which in our case corresponds to the quotient symmetry $\tilde{U}(1)$) while the topology in the SPT phase we are studying is associated with the dual 1-form symmetry.     

In the decomposition of $\varphi$ above, $n$ and $m$ label the twisted sector, and $\delta \varphi$ is neutral under the 1-form symmetry generated by $W$. Nevertheless, we may locally deform $\delta \varphi$  to some separate $\tilde{\theta}$-domain walls where locally its value jumps by $\tilde{\theta}$. Similar to the 1-D case and the discrete case, by assuming fusion linearity, we may \textit{formally} assign a charge $\tilde{\theta}/\pi$ for $0\le \tilde{\theta} \le \pi$ under the dual 1-form symmetry to a domain wall where $\delta \varphi$ changes by $\tilde{\theta}$.  This is a manifestation of the mixed anomaly and the symmetry fractionalization.  In the superfluid/Higgs phase, $\delta \varphi$ is small such that the winding numbers of $\varphi$ is conserved. Proliferation of the winding, and equivalently the inserted fluxes, breaks the $\hat{\bbz}_2$ 1-form symmetry and recovers the $\tilde{U}(1)$ simultaneously, leading to a SET phase.

\subsection{SET phase}
Having discussed the SPT phase, we now briefly touch upon the emergent $\tilde{U}(1)$-SET phase [Fig.~\ref{fig2d}(b)].

The SSB of a discrete higher-form symmetry leads to a topologically ordered phase \cite{gaiotto2015generalized, mcgreevy2023generalized}, a phase with long-range entanglement and ground state degeneracy depending on the topology of the base space.  The SSB of the dual $\hat{\bbz}_2$ 1-form symmetry leads to a $\bbz_2$-topological ordered phase whose excitations are the same as Kitaev's toric code or (untwisted) quantum double model \cite{Kitaev2003}. When OBC is used, the boundaries can be gapped \cite{bravyi1998quantum, kitaev2012models}. This phase is also enriched by the $\tilde{U}(1)$ because the quotient $\tilde{U}(1)$ symmetry is preserved.

The topological charges are gauge charges $e$, $\pi$-fluxes $m$, and their bound state $em$. The last two types are not dynamical due to the exactness of the 1-form symmetry, or equivalently the zero-flux condition. Since $\tilde{U}(1)$ does not commute anyon types, there is no obstruction to the symmetry fractionalization \cite{barkeshli2019symmetry}. Hence, it is classified by $[w] \in H^2(\tilde{U}(1), \Aa)$ where $\Aa$ is the finite group whose elements are the
Abelian topological charges of the unitary modular tensor category $\Ca$ with group multiplications
given by their corresponding fusion rules \cite{barkeshli2019symmetry}. In this case, $\Aa = \bbz_2 \times \bbz_2$, so  $ H^2(\tilde{U}(1), \Aa) = \bbz_2 \times \bbz_2$. Explicitly, a representative cocycle is given by $ 
w(\theta, \theta') = m^{\lfloor \frac{\theta +\theta'}{2\pi}\rfloor},
$
where $\theta, \theta' \in [0, 2\pi)$ parametrizes $\tilde{U}(1)$ and $\lfloor \cdot \rfloor$ is the floor function.  The nontriviality of 
 $w$, as a manifestation of the mixed anomaly we discussed above, dictates the fractionalization of charges under $\tilde{U}(1)$.  Alternatively, the  system can be viewed as living on the surface of an SPT phase in 3-D protected by $\tilde{U}(1)$ and $\hat{\bbz}_2$. To trivially gap out the system, $\tilde{U}(1)$ has to be broken.

\subsection{Generalized DQCP}
As we have discussed above, even though $\pi$-vortices are suppressed once the zero-flux condition is enforced, the proliferation of topological phase winding excitations drives the SPT phase to the SET phase. Since the original superfluid-insulator transition is continuous and only a finite subgroup is gauged, we expect this inherited transition to be continuous as well \cite{su2023boundary}. Thus, we obtain a generalized DQCP from the gapless SPT phase with preserved $\hat{\bbz}_2$ 1-form symmetry and broken $\tilde{U}(1)$ to a SET phase with $\tilde{U}(1)$ [see Fig.~\ref{fig2d}(b)].

As a result of partial gauging of a finite group, much of the information encoded in the order/disorder parameters can be directly read from the original superfluid-insulator transition. The critical point can be determined by the change in the scaling laws of different order/disorder parameters.  The symmetry breaking of higher-form symmetries at critical points has been investigated in the recent literature \cite{Wu2021categorical, wu2021universal, wang2022scaling}.  If all symmetries involved including the higher-form symmetry are preserved at the critical point in some systems, we may get a 2-D analog of the intrinsically gapless SPT in 1-D by invoking similar anomaly arguments.  

This generalized DQCP is essentially the same as the so-called $XY^*$  transition obtained from the conventional 3-D $XY$ critical point \cite{isakov2012universal, Swingle2012}. The difference is that it is $\psi \sim bb$ rather than $b$ that is treated as the fundamental degree of freedom \cite{isakov2012universal}. At the $XY^*$  transition, $b$ undergoes an ordinary $XY$ transition. Since $\psi$ is a composite operator of two $b$'s, the power law scaling exponent of $\langle \psi^\dagger_i \psi_j\rangle$ gets significantly modified $\eta_{bb} \sim 1.49$ from  $\eta_b \sim 0.03$ for $b$. The divergence of the correlation length $\nu \sim 0.67$ and the isotropy of the space and time dimensions $z \sim 1$ were verified to be the same as in the conventional 3-D $XY$ universality class. These statements were verified numerically \cite{isakov2012universal}. 

In terms of entanglement entropy $S_R$ of a smooth simply connected region $R$ without corners, it is known that, other than the leading perimeter law term $S_A \propto l \equiv |\partial R|$,  there is a logarithmic subleading correction in a SSB phase of a continuous symmetry \cite{metlitski2011entanglement} and a topological subleading correction in a topologically ordered phase \cite{kitaev2006, levin2006}. At the critical point, $S_A$ takes the form of $ 
S_A = \alpha l - \beta$ with $\beta = \beta_{XY} + \beta_{\bbz_2}
$. Here $\beta_{XY}$ comes from the ordinary $XY$ transition, i.e., SSB of $U(1)$,  and $\beta_{\bbz_2} = \ln 2$ is the topological entanglement entropy of the $\bbz_2$ topologically ordered phase \cite{Swingle2012}.

If we consider OBC, there is also a generalized boundary phase transition between the gapless SPT/Higgs phase and the $\tilde{U}(1)$-enriched topological phase. It would be interesting to investigate this boundary phase transition.

\subsection{Perturbations}
In the above discussion, we have imposed the zero-flux condition by taking $g \to 0$ in order to preserve the exactness of the dual 1-form. The existence of the $\sum \sigma^x_i$ term in Eq.~(\ref{eq2d}) explicitly breaks this symmetry. However, we expect the perturbation changes neither the topological order in the SET phase nor the gapless boundary modes in the SPT phase in 2-D due to the robustness of higer-form symmetries \cite{mcgreevy2023generalized}. In particular, in the original Fradkin-Shenker phase diagram for 2-D Ising gauge theory coupled to matter with PBC, both the deconfined phase, the Higgs phase and the transition in between are robust with the introduction of the small polarizing field. 

To study the edge physics,  the authors in Ref.~\cite{verresen2022higgs} numerically demonstrated the robustness of the topological edge modes in the SPT/Higgs phase. On the other hand, the robustness of $\bbz_2$ topological order with open boundaries under perturbation is also numerically investigated in Ref.~\cite{robustness}. Based on their results, the robustness depends both on the boundary type (rough or smooth) and the perturbation type ($\sigma^x$ or $\sigma^z$). For rough boundaries [see Fig.~\ref{fig:PW}(a)] and perturbations of the form  $\sigma^x$,  the topological order is robust.  It is natural to expect resilience in the edge physics. 

Based on the robustness of both phases, it is also natural to expect the SPT/Higgs-SET transition to remain continuous and robust, even with small perturbations that explicitly break the 1-form symmetry. In this case, we can regard the 1-form symmetry to be emergent \cite{pace2023}, and the properties of the generalized DQCP should remain intact. In this sense, generalized DQCPs can be a generic type of quantum criticalities and deserve to be investigated in more details in future work.

% ------------------------------------------------------------------------------
\subsection{Experimental realizations}
\label{sec:exp}
The Bose-Hubbard model in 1-D and higher dimensions has been realized in such systems as cold atoms on optical lattices \cite{jaksch1998, greiner2002quantum, Stoferle2004, spielman2007, Bloch2008, browaeys2020many}, and the continuous superfluid-insulator transition has been observed. Lattice gauge theories have also been simulated in such systems \cite{banuls2020simulating,wiese2013ultracold,tagliacozzo2013simulation}. Recently, the $\bbz_2$ topological order has been realized and measured in Rydberg atoms on a 2-D Ruby lattice \cite{verresen2021prediction, semeghini2021probing}. It is more complicated to simulate gauged matter theories, but there are also some recent experimental progress in this direction. For example, a similar $\bbz_2$-gauge Bose-Hubbard model in 2-D was studied in Ref.~\cite{homeier2023realistic} with the idea of realizing the gauge constraints by using simplified local pseudogenerators \cite{halimeh2022}. We believe quantum simulation with cold atoms is a promising platform to realize the unconventional phases and  quantum criticalities proposed in this work.

Realizations of the gapless SPT phases and the generalized DQCPs discussed in our work may be also possible in other solid state systems. For example, 
since the generalized DQCP is essentially the $XY^*$ transition studied before  
\cite{isakov2011topological, isakov2012universal}, we may start with an ordinary Bose-Hubbard system with fractionalized excitations. As long as the low energy effective theory is described by an emergent $\bbz_2$-gauged Bose-Hubbard model, we may test the analysis in our work. 

A recent trend in the past few years has been realizing the gauging process by using finite-depth unitaries, measurement, and feedforward so topological ordered states can be obtained efficiently \cite{tantivasadakarn2021long, Tantivasadakarn2023, iqbal2023topological, foss2023experimental}. Other phases such as Higgs phases \cite{Kuno2023production} and continuous symmetry breaking states \cite{hauser2023continuous}, and phase transitions \cite{negari2023measurement, kuno2023bulk} have been proposed. Some SET phases can also be obtained by partial gauging \cite{li2023symmetry}. Realizing a SET phase by partially gauging a $U(1)$ symmetry is also very natural. To simulate and study a gapless SPT phase and the generalized DQCP discussed in this work in these adaptive circuits is a fascinating direction.

% ------------------------------------------------------------------------------
\section{Conclusions}
\label{sec:concl}
In this work, we investigated the emergent $\bbz_2$-gauged matter theory of the 1-D and 2-D Bose-Hubbard model coupled to Ising degrees of freedom. We analyzed the inherited phase diagram from that of the ungauged superfluid-insulator version. In 1-D, we identified the superfluid phase to be an intrinsically gapless SPT phase protected by $W\times U(1)$, $W$ being the Ising spin reflection symmetry. In the low energy theory, $W$ can be viewed as the dual $\hat{\bbz}_2$ symmetry.  We discussed the effective action which includes a mixed anomaly term between  $W$ and the quotient $\tilde{U}(1)$ symmetry $\omega$, and a topological term $\alpha$ dictating the edge degrees of freedom if there is an open boundary.  The 't Hooft anomaly in $\alpha$ is matched by that in $\omega$. We argued that the gapless SPT phase is described by a $\bbz_2$-gauged compact boson CFT, which is also supported by DMRG computations.   In 2-D, we focused on the zero-flux limit and concluded, by adapting the  ``Higgs = SPT" argument, that the superfluid phase is also a gapless SPT protected by higher form symmetries whose gaplessness comes from the Goldstone modes due to the SSB of the quotient $\tilde{U}(1)$. We studied the excitations, especially the Goldstone modes whose winding number is related to the charge of the $\tilde{U}(1)$ domain wall under the dual $\hat{\bbz}_2$ 1-form symmetry $W$, which is a direct manifestation of the mixed anomaly between the two symmetries. The other phase is the insulating phase corresponding to the SSB of the $W$ with $\tilde{U}(1)$ preserved, i.e., $\tilde{U}(1)$-enriched $\bbz_2$ topological order. Then we analyzed the transition between the gapless SPT/Higgs phase and the    $\tilde{U}(1)$-SET phase, which is a generalized DQCP. The robustness of the gapless SPT/Higgs phase, the SET phase, and the generalized DQCP between them toward perturbations that explicitly break the $\hat{\bbz}_2$ 1-form symmetry is discussed. Possible experimental realizations using quantum simulations with cold atoms are also proposed. 

The idea of partially gauging a finite subgroup discussed in this work is  straightforward and general.  In principle, we can start with any system that has a SSB of a generic continuous symmetry, including generalized symmetries, and then perform the partial gauging to arrive at novel phases and phase transitions in between. The system can even be topological at the outset. Extension to higher dimensions is straightforward.  The colluding roles of the topological term $\alpha$ and the emergent anomaly $\omega$  in constructing general (intrinsically) gapless SPT  phase deserve further elaboration. It would also be interesting to study deformations of the gapless SPT phases  away from their fixed-point so that the edge localization  length $\xi_e$ is not strictly zero. 

As we have mentioned earlier on, our analysis generalizes easily to fermionic systems. In particular, the 1-D intrinsically gapless SPT phase works for a free fermion gas or a Luttinger liquid \cite{borlar2021gauging} and is generalizable to generic critical point or a Fermi liquid in higher dimensions. Generically, introduction of a weakly fluctuating gauge field may destabilize the system and drive the system to other phases, such as superconductors. It would be interesting to construct such a stable  intrinsically gapless fermionic SPT phase. A good starting point may be exactly solvable models of free lattice fermions coupled to Ising spins on the link. When it is not analytically solvable, numerical simulations using, e.g., the determinant quantum Monte Carlo method similar to      
Refs.~\cite{gazit2017emergent, gazit2018confinement} can give us more valuable insights.   We hope our work can stimulate more endeavors along these directions.

\section{Acknowledgement} L.S. and M.Z. would like to thank Yi-Zhuang You for insightful discussions. L.S. thanks Ivar Martin for useful discussions. L.S. is supported by the Simons Foundation (Grant No. 669487). M.Z. would like to thank Ryan Thorngren for conversations. M.Z. would also like to thank the Institute of Advanced Study at Tsinghua University China for hospitality where part of this work was done.  M.Z. is supported by NSF Grant No. DMR-2238360.

\appendix

\section{Emergent $\bbz_2$-gauged $q$-state clock model}
\label{app_a}
In the main text, we focused on the case where the $\bbz_2$ subgroup of the continuous $U(1)$ is effectively gauged. Some of the properties we discussed there can already be found when the total symmetry group is discrete. In this appendix, we present similar analysis of the $q$-state clock model with discrete on-site $\bbz_q$ symmetry. We focus on even $q$ cases, so that there exists a $\bbz_2$ subgroup that can be subsequently gauged. Both 1-D and 2-D cases are discussed.
\subsection{1-D}
For the Ising model, i.e., when $q=2$, gauging the $\bbz_2$ is equivalent to a Kramers-Wannier transformation from the original Ising model to the dual Ising model with a dual 0-form $\hat{\bbz}_2$ symmetry. The minimal nontrivial case corresponds to $q=4$, which is also discussed in Refs. \cite{zhang2023, su2023boundary}. The $\bbz_2$-gauged 4-state clock model has global symmetry $\tilde{\bbz}_2\times\hat{\bbz}_2$ with a mixed anomaly between the two symmetries characterized by the non-trivial extension class in $H^2(\tilde{\bbz}_2,\hat{\bbz}_2)=\bbz_2$.
A  concrete lattice model can be written down as follows,
\begin{equation}
\begin{split}
    H=&-J\sum_j\left(C_j^\dagger \tau^z_{j+1/2} C_{j+1} + h.c.\right)- h\sum_j \left(S_j+S_j^\dagger\right)\\
    &- K\sum_j \tau_{j-1/2}^xS_j^2\tau_{j+1/2}^x,
    \end{split}
\end{equation}
where $C^4_j=S^4_j=1$ and $C_jS_j=\text{e}^{i\frac{2\pi}{4}}S_jC_j$. We fix $J =1$. Similar to the case in the main text, we have a minimal coupling between clock degrees of freedom on sites and Ising spins on bonds, where the large $K$ limit effectively implements the gauging with the gauge constraint given by $G_j=\tau_{j-1/2}^xS_j^2\tau_{j+1/2}^x=1$. The global symmetry is $V\times W$, with 
\begin{equation}
    V=\prod_j S_j,\quad W=\prod_j\tau_{j+1/2}^z,
\end{equation}
where at low energy $W$ becomes the dual $\hat{\bbz}_2$ symmetry and $V$ is the quotient symmetry. There is a subtlety for the $V$ symmetry when there is an open boundary. For periodic boundary condition (PBC), we have $V^2=\prod_jS_j^2=\prod_j\tau_{j-1/2}^x\tau_{j+1/2}^x=1$, by using the low energy gauge constraint, making $V$ explicitly $\bbz_2$. However, for open boundary condition (OBC), we have instead the nontrivial identity $V^2=\tau_{1/2}^x\tau_{L+1/2}^x$. Here $L$ is the number of sites in the open chain. 

In the case of $h\gg 1$, $S_j$ will be polarized, which implies that $\langle \tau_{m-1/2}^x\tau_{n+1/2}^x \rangle=\prod_{m\leq j\leq n}S_j^2\neq 0$, i.e., there is long range order in $\tau^x$, leading to SSB of $W$ while $V$ is preserved. On the other hand, when $h\ll 1$, there will be SSB in $V$ but with $W$ preserved. The SSB in $V$ directly inherits from the SSB of the ungauged clock model, since the $\bbz_2$ gauging corresponds to summing over twisted boundary conditions, which does not change the long-range correlation of the order parameter \cite{,zhang2023, su2023boundary}. Due to the SSB of $V$, which is simply $\tilde{\bbz}_2$ in the bulk, the ground state is 2-fold degenerate $|\psi_1\rangle$ and $|\psi_2\rangle$ with $V|\psi_1\rangle=|\psi_2\rangle$. 

Furthermore, in the $V$ SSB phase, for either of the degenerate ground state $|\psi_\alpha\rangle$, there is non-trivial string order parameter given by $\langle C_m^\dagger \tau_{m+1/2}^z...\tau_{n-1/2}^zC_n\rangle\neq 0$, signifying that the $V$ SSB phase is in fact the Higgs/SPT phase \cite{verresen2022higgs}. To see the non-trivial edge states, we consider a semi-infinite chain with one open boundary at site $L$. Then we have $V^2=\tau_{L+1/2}^x$, so that $V^2W=-WV^2$. For a ground state $|\psi_\alpha\rangle$, both $V^2$ and $W$ are symmetries. Due to the anti-commutation of the two symmetry operators, we can similarly argue that $V^2|\psi_\alpha\rangle$ and $|\psi_\alpha\rangle$ are degenerate and the degeneracy comes from the edge since $V^2$ is localized at the edge. This way, we explicitly see the edge degeneracy for each of the bulk degenerate ground states. 

The analog of the intrinsically gapless SPT of the $U(1)$ case shows up when $q\geq 5$. It is known that the 1-D quantum clock model without gauging has two critical points, both are of BKT type and dual to each other, and there is a critical phase with emergent $U(1)$ symmetry in between the two critical points \cite{ortiz2012dualities}. The minimal non-trivial case with mixed anomaly after gauging is the 8-state clock model. After gauging, the emergent symmetry is $\tilde{\bbz}_4\times\hat{\bbz}_2$, where there is a mixed anomaly characterized by the non-trivial extension class in $H^2(\tilde{\bbz}_4,\hat{\bbz}_2)=\bbz_2$.

\begin{figure}[tb]
\centering
\includegraphics[width=0.7\columnwidth]{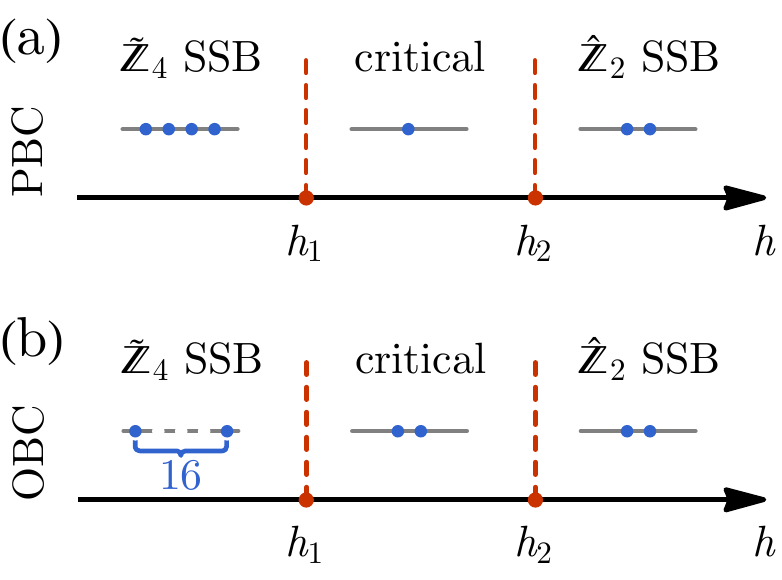} 
\caption{Ground state degeneracy in different phases of the 1-D $\bbz_8$-clock model with its $\bbz_2$ subgroup gauged, under PBC (a) and under OBC (b).}

\label{fig:Z8}
\end{figure}

Fig.~\ref{fig:Z8} shows the schematic phase diagram with the corresponding ground state degeneracies under both PBC and OBC. In the large $h$ limit, the dual $\hat{\bbz}_2$ symmetry is spontaneously broken, which is labeled as the $\hat{\bbz}_2$ SSB phase.  In the small $h $ limit, we have the $\tilde{\bbz}_4$ SSB phase. In the intermediate coupling regime ($h_1<h<h_2$), the system is in a critical phase, where both of the two symmetries are preserved.  There is an emergent symmetry $U(1) \times U(1) \supset \tilde{\bbz}_4\times\tilde{\bbz}_2$ and the critical phase is described by the ($\bbz_2$-gauged) compact boson CFT. Indeed, this phase is an analog of the intrinsically gapless SPT phase we discussed in the main text.

\subsection{2-D}
For 2-D $q$-state clock model, the Hamiltonian takes similar form with that of the $\bbz_2$-gauged Bose-Hubbard model described in the main text,
\begin{equation}
\begin{split}
    H=&-J\sum_{i,j\in \partial e}\left(C_i^\dagger \tau_e^zC_j + h.c.\right)-h\sum_j\left(S_j+S_j^\dagger\right)\\
    &-\frac{1}{g}\sum_p\prod_{e\in \partial p}\tau^z_e-g\sum_e\tau_e^x\\
    &-K\sum_jS_j^{q/2}\prod_{e,j\in \partial e}\tau_e^x,
\end{split}
\end{equation}
where $C_j^q=S_j^q=1$ and $C_jS_j=\text{e}^{i\frac{2\pi}{q}}S_jC_j$. We fix $J =1$. In the zero flux limit $g\to 0$ and large $K$ limit, the global symmetries are the quotient 0-form $\tilde{\bbz}_{q/2}$ and the 1-form $\hat{\bbz}_2$, given by,
\begin{equation}
    V=\prod_j S_j,\quad W=\prod_{e\in \gamma}\tau^z_e,
\end{equation}
 where $\gamma$ is a loop running along the bonds of the lattice. Notice that in order for there to be an emergent mixed anomaly between $V$ and $W$, the group extension of $V$ by $W$, classified by $H^2(\tilde{\bbz}_{q/2},\hat{\bbz}_2)=\bbz_{\text{gcd}(q/2,2)}$ has to be non-trivial, meaning that $q/2$ has to be even, i.e., $q$ is an integer multiple of 4. 
Similar to the 1-D case, due to the mixed anomaly between $V$ and $W$, $W$ is SSB while $V$ is preserved when $h \gg 1$. This is the $\bbz_2$ topological order enriched by $\tilde{\bbz}_{q/2}$, classified by $H^2(\tilde{\bbz}_{q/2},\bbz_2\times\bbz_2)=\bbz_{\text{gcd}(q/2,2)} \times \bbz_{\text{gcd}(q/2,2)}$. 

When $h \ll 1$, we have SSB for $V$ with $W$ preserved. Therefore, the bulk is $q/2$-fold degenerate, and each degenerate ground state has additional degenerate edge states when there is an open boundary. The argument is the same by considering the two anti-commuting symmetries $V^{q/2}$ and $W$ of the ground states. Therefore, this is again the Higgs/SPT phase. The $\tilde{\bbz}_{q/2}$-enriched topological order and the $\hat{\bbz}_2$ 1-form protected SPT with $q/2$-fold bulk degeneracy are separated by a generalized DQCP. 

\section{Elimination of gauge constraints}
 \label{app_eli}
In the main text, based on the mixed anomaly between the quotient $\tilde{U}(1)$ symmetry and the dual 1-form $\hat{\bbz}_2$ symmetry in the emergent gauge theory, we claimed that these two symmetries cannot both be on-site. This can be demonstrated easily by eliminating the gauge constraints. The elimination may be achieved by performing a unitary transformation consisting of controlled gates and Hadamard transformations   \cite{Yoshida2016top}. Here we follow Ref.~\cite{su2023boundary} to gain more intuition.

After gauging, i.e., implementing the gauge condition $\prod_{e, i\in\partial e}\sigma^x_e=(-1)^{n_i}$, the onsite boson states are divided into the boson parity even sector and the parity odd sector, depending on the sign of the star operator $\prod_{e, i\in\partial e}\sigma^x_e$. Therefore, the new onsite boson basis can be denoted as $|1,\tilde{n}\rangle$ and $|-1,\tilde{n}\rangle$, where the first number labels the onsite boson parity, determined by $\prod_{e, i\in\partial e}\sigma^x_e$, and the second number labels the new local boson states with $\tilde{n}=0,1,2,...$ in the corresponding sector. Notice that we have a one-to-one correspondence between the new basis and the original ungauged basis, given by $|1,\tilde{n}\rangle\leftrightarrow |2\tilde{n}\rangle$ and $|-1,\tilde{n}\rangle\leftrightarrow |2\tilde{n}+1\rangle$. Therefore, no degrees of freedom are lost, as it should be. Expressed in the new basis, the boson number operator becomes
\begin{equation}
\begin{split}
    \hat{n}_i\to &\frac{1+\prod_{e,i\in\partial e}\sigma^x_e}{2}2\hat{\tilde{n}}_i+\frac{1-\prod_{e,i\in\partial e}\sigma^x_e}{2}\left(2\hat{\tilde{n}}_i+1\right)\\
    &=2\hat{\tilde{n}}_i+\frac{1-\prod_{e,i\in\partial e}\sigma^x_e}{2}.
\end{split}
\label{eq:b1}
\end{equation}
The action of the boson creation/annihilation operator should be accompanied with a flip in $\prod_{e, i\in\partial e}\sigma^x_e$ since $(-1)^{n_i}$ changes sign. Consider the gauge invariant minimal coupling term $b_i^\dagger \sigma_{e_{ij}}^z b_j$.
In the new basis, 
\be 
  b_i^\dagger \sigma_{e_{ij}}^z b_j\to A_i \sigma^z_{e_{ij}} B_j.
\label{eq:b2}
\ee 
where 
\begin{equation}
\begin{split}
    & A_i \equiv  \frac{1+\prod_{e,i\in\partial e}\sigma^x_e}{2}+\frac{1-\prod_{e,i\in\partial e}\sigma^x_e}{2}\tilde{b}_i^\dagger,\\
    & B_i\equiv  \frac{1+\prod_{e,i\in\partial e}\sigma^x_e}{2}\tilde{b}_i+\frac{1-\prod_{e,i\in\partial e}\sigma^x_e}{2}.
\end{split}
\end{equation} 
In the new basis, the action of the dual 1-form $\hat{\bbz}_2$ symmetry remains unchanged as $W=\prod_{e\in\gamma}\sigma^z_e$ for closed loop $\gamma$, but the $\tilde{U}(1)$ symmetry is now implemented by 
\begin{equation}
    \tilde{X}(\theta)=\prod_i\mathrm{exp}\left[i\left(2\hat{\tilde{n}}_i+\frac{1-\prod_{e,i\in\partial e}\sigma^x_e}{2}\right)\frac{\theta}{2}\right],
\end{equation}
which is explicitly non-onsite. It is easy to see that $ \tilde{X}(\theta) =  \tilde{X}(\theta+2 \pi)$ for PBC. Combining Eq.~(\ref{eq:b1}) and (\ref{eq:b2}), we can obtain the Hamiltonian of the Bose-Hubbard model (with the zero-flux condition) expressed in the new basis where the gauge constraints have already been encoded. 

Note that the quantity $\hat{\tilde{N}}  =\sum_i \hat{\tilde{n}}_i$ itself is not conserved. Instead, the term $(1-\prod_{e,i\in\partial e}\sigma^x_e)/2$  contributes a fractional charge 1/2 to $\hat{\tilde{N}}$. By invoking the ``electromagnetic duality" for the $\bbz_2$ gauge field: $\sigma^x \leftrightarrow \sigma^z$, we can regard the zero-flux condition $\prod_{e \in \partial p} \sigma_e^z   = 1$ as the new ``Gauss law" for the ``gauge field" $\sigma^x$ while  $\prod_{e,i\in\partial e}\sigma^x_e$ as the new ``magnetic" flux operator. Thus, the new magnetic flux carries a fractional charge under $\hat{\tilde{N}}$, a manifestation of the mixed anomaly. Since the anomalous system may be viewed as a boundary of an SPT phase in the (3+1)-D bulk, as mentioned in Ref.~\cite{su2023boundary}, it is natural to envision a discrete realization of the bulk by decorating the magnetic monopoles of the $\hat{\bbz}_2$ gauge field with unit charges under $\tilde{U}(1)$, which is an analog of the continuum construction in Ref.~\cite{jian2021physics}.

\section{Grand canonical ensemble in 1-D}
\label{app_gce}
\begin{figure}[tb]
\centering
\includegraphics[width=0.98\columnwidth]{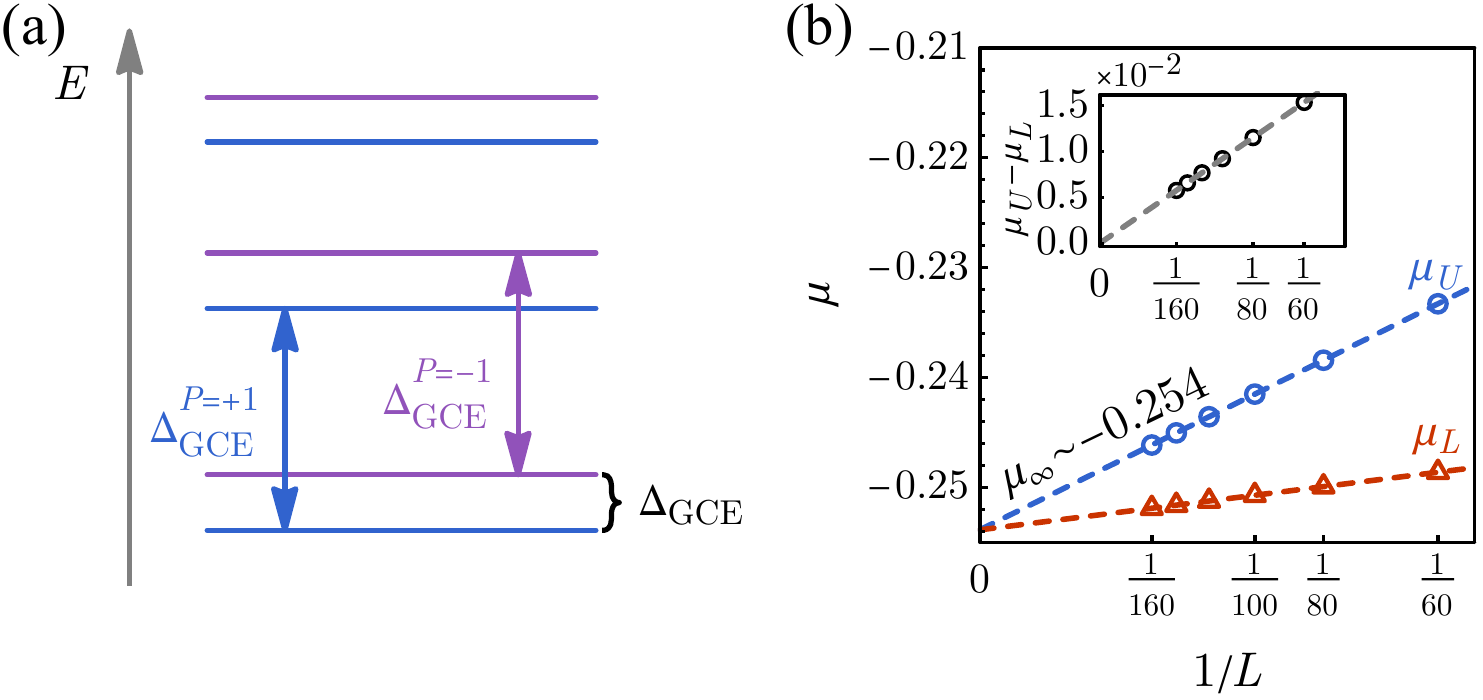} 
\caption{(a) Schematic diagram for the energy levels in the grand canonical ensemble, where parity even and parity odd states coexist. $\Delta_{\text{GCE}}^{P = \pm 1}$ is the gap in the parity even/odd sector and $\Delta_{\mathrm{GCE}}$ is the true gap above the doubly degenerate ground state. (b) The upper bound $\mu_U$ and the lower bound $\mu_L$ for the chemical potential in order to have unit filling at different system sizes. The inset shows the difference between the two bounds. The parameters used are $t=0.5$ and $U=1.0$.}
\label{fig:fig7}
\end{figure}

In Sec.~\ref{sec1d}, we used canonical enemble (CE) in 1-D for simplicity since the particle number conservation cannot be violated due to the Mermin-Wagner theorem.  
In this appendix, we support  this statement by presenting some DMRG results  for the $\bbz_2$-gauged Bose-Hubbard model in the context of grand canonical ensemble (GCE) where the total particle number can vary.  GCE is more general and it allows energy levels consisting of both even and odd parity states [see Fig.~\ref{fig:fig7}(a)]. The existence of states with different parities is a consequence of the boundaries when using OBC, where the parity operator $P=\sigma_{1/2}\sigma_{L+1/2}$ can still take values $\pm 1$. However, GCE is computationally more challenging in DMRG since particle number is not fixed. On the other hand, we observed numerically that as long as the chemical potential $\mu$ is carefully tuned to ensure unit filling for the GCE ground state, both the ground state and the next excited state have no fluctuations in total boson number, i.e., they have fixed boson number. A nice consequence of this observation is that the GCE low energy states can now be related to those of the CE, which renders the numerical calculation easier and more tractable for larger system sizes. We can still show that the superfluid phase is a gapless SPT phase with double degeneracy.

The GCE Hamiltonian $H_{\mathrm{GCE}}(\mu,\hat{N})$ is related to the CE Hamiltonian $H_{\mathrm{CE}}(\hat{N})$ as the following,
\begin{equation}
    H_{\mathrm{GCE}}(\mu,\hat{N})=H_{\mathrm{CE}}(\hat{N})-\mu \hat{N},
\end{equation}
where $\hat{N}$ is the total boson number operator. As mentioned previously, the chemical potential $\mu$ can be tuned to achieve a ground state with unit filling and fixed boson number, i.e., $\langle\hat{N}\rangle=L$ in the ground state. Here $L$ is the number of boson sites which for simplicity is taken to be even. Notice, however, that the proper $\mu$ has a strong size dependence. To carry out a finite size scaling of the gap, we extract the thermodynamical information by bounding the gap as follows.  

For a typical set of parameters both the gap in the parity even sector $\Delta_{\text{GCE}}^{P = +1}$ and the gap in the parity odd sector $\Delta_{\text{GCE}}^{P = -1}$ are much larger than the true gap $\Delta_{\text{GCE}}$ which is between a parity even state and a parity odd state [Fig.~\ref{fig:fig7}(a)]. Note that the ground state is exactly doubly degenerate as in the CE. The GCE ground state energy is given by,
\begin{equation}
    E_{\mathrm{GCE}}(\mu,L)=E_{\mathrm{CE}}(L)-\mu L,
\end{equation}
and satisfies the following conditions,
\be  
     E_{\mathrm{GCE}}(\mu,L)\leq E_{\mathrm{GCE}}(\mu,L\pm 1),
\ee
which implies
\begin{equation}
\begin{split}
    \mu_L &\equiv E_{\mathrm{CE}}(L)-E_{\mathrm{CE}}(L-1)\\
    &\leq \mu \leq E_{\mathrm{CE}}(L+1)-E_{\mathrm{CE}}(L)\equiv \mu_U.
    \label{eq:mu-bounds}
\end{split}
\end{equation}
Fig.~\ref{fig:fig7}(b) shows the system size dependence of the lower and upper bounds of the chemical potential based on Eq.~(\ref{eq:mu-bounds}). The two bounds converge to $\mu_\infty \approx -0.254$ in the thermodynamic limit, which is also reflected in the inset showing the difference between the bounds approaching 0. 

Both $\Delta_{\text{GCE}}^{P = \pm 1}$ decay to zero following a power law $1/L$. Furthermore, the true gap is given by 
\begin{equation}
\begin{split}
    \Delta_{\mathrm{GCE}}&=\mathrm{min}\{E_{\mathrm{GCE}}(\mu,L+1)-E_{\mathrm{GCE}}(\mu,L),\\
    & \quad \quad \quad \quad E_{\mathrm{GCE}}(\mu,L-1)-E_{\mathrm{GCE}}(\mu,L)\}\\
    &=\mathrm{min}\{\mu_U-\mu,\mu-\mu_L\}\\
    &\leq \mu_U-\mu_L.
\end{split}
\end{equation}
Since $\mu_U-\mu_L$ is shown to decay to 0 in the thermodynamic limit as $1/L$ (see inset of Fig.~\ref{fig:fig7}(b)), the bulk gap $\Delta_{\mathrm{GCE}}$ also decays to 0 at least as fast as $1/L$.

\section{Anomaly action}
\label{app_ano}
In this appendix, we provide more details regarding the 't Hooft anomalies that show up in the main text. We will emphasize the comparison between the mixed anomaly from gauging a finite subgroup \cite{tachikawa2020} and the emergent anomaly through separation of gapped and gapless degrees of freedom \cite{Thorngren2021}.

Consider the following central extension
\be 
    1\to \bbz_2 \xrightarrow{i} U(1) \xrightarrow{\pi} \tilde{U}(1) \to 1.
\ee
Here $i$ is inclusion and $\pi$ is projection. The extension corresponds to the nontrivial element $e$ in $ H^2(\tilde{U}(1), \bbz_2) =\bbz_2$. To Let $G = U(1)$ be the total global symmetry free from anomalies. Then we can turn on a flat background gauge field $A^{U(1)}$ on a closed (1+1)-D spacetime $M$.
Then 
\be 
A^{U(1)}= i(A^{\bbz_2}) + r(A^{\tilde{U}(1)}),
\ee 
where $r$ lifts $A^{\tilde{U}(1)}$ into $A^{U(1)}$ and satisfies $\pi(r) = \text{Id}$. The flatness of $A^{U(1)}$ implies that the $A^{\bbz_2}$ sees the flux of $A^{\tilde{U}(1)}$:
\be 
dA^{\bbz_2} = e(A^{\tilde{U}(1)}) = \frac{d A^{\tilde{U}(1)}}{2 \pi}\ \text{mod 2},
\label{eq_AA}
\ee
where we have omitted $r$ for simplicity. We can gauge the $\bbz_2$ subgroup of $U(1)$ by making  $A^{\bbz_2}$ dynamical which we will denote as $a^{\bbz_2}$. 
Then
\be 
da^{\bbz_2} = e(A^{\tilde{U}(1)}) = \frac{d A^{\tilde{U}(1)}}{2 \pi}\ \text{mod 2}.
\label{eq_da}
\ee
After gauging $\bbz_2$, there is a dual quantum symmetry $\hat{\bbz}_2$ showing up \cite{vafa1986modular}. It can again be coupled to its background field $A^{\hat{\bbz}_2}$ as follows
\be S = \pi i \int_M A ^{\hat{\bbz}_2} \cup a^{\bbz_2}.
\label{eq_zz}
\ee
Since $a^{\bbz_2}$ is not closed when $A^{\tilde{U}(1)}$ is nontrivial, as a result of Eq.~(\ref{eq_da}), there is an mixed anomaly between the dual symmetry $\hat{\bbz}_2$ and quotient $\tilde{U}(1)$ which is characterized by an anomaly action in a (2+1)-D dimensional bulk $Y$
\be 
\omega = \frac{1}{2} A ^{\hat{\bbz}_2} \cup d a^{\bbz_2} =  \frac{1}{2} A^{\hat{\bbz}_2} \cup \frac{d A^{\tilde{U}(1)}}{2 \pi},
\label{eq_omeg}
\ee 
where $A^{\hat{\bbz}_2}$ and $A^{\tilde{U}(1)}$ are extended to the bulk $Y$. When $M = \partial Y$, the anomaly is canceled. In the main text, the gauge theory is emergent, so the mixed anomaly is also emergent in the low energy theory. 

In Ref.~\cite{Thorngren2021}, the $\bbz_2$ group is not gauged but gapped out by interactions in the sense that $\bbz_2$ only acts nontrivially on the gapped degrees of freedom  while $\tilde{U}(1) $ is the symmetry that acts nontrivially on the low energy degrees of freedom. Turning on the background fields, we arrive at Eq.~(\ref{eq_AA}) as well. If the symmetry  $\tilde{U}(1)$ acting on the gapless degrees of freedom has an emergent anomaly, it is possible to construct a gapless SPT phase. 

In particular, if $U(1)$ is broken to $\bbz_4$, then since $H^2(\bbz_4, U(1)) =0$, there is no gapped SPT phase in 1-D. However, an intrinsically gapless SPT phase can exist when there is an emergent anomaly, which is captured by 
\be 
\omega = \frac{1}{2} A^{\bbz_2} \cup  d A^{\bbz_2} 
\ee
in the higher-dimensional bulk $Y$. Indeed, since the total symmetry group $G = \bbz_4$ is anomaly-free, the low energy anomaly $\omega (A^{\bbz_2})$ must be compensated by a counterterm $\alpha(A^{\bbz_2}, A^{\tilde{\bbz}_2})$ satisfying the anomaly vanishing equation \cite{Thorngren2021}
\be 
\omega(A^{\bbz_2}) = d \alpha(A^{\bbz_2}, A^{\tilde{\bbz}_2}).
\ee  
Here $A^{\tilde{\bbz}_2}$ is the background field of the quotient symmetry $\tilde{\bbz}_2 \equiv \bbz_4/\bbz_2$ acting on the gapped degrees of freedom. The partition function then may be written as  
\be Z = e^{2\pi i \int_Y \omega(A^{\bbz_2}) } e ^{-2\pi i \int_M \alpha(A^{\bbz_2}, A^{\tilde{\bbz}_2})}. \ee 
One solution to the anomaly vanishing equation is given by $\alpha(A^{\bbz_2}, A^{\tilde{\bbz}_2}) =    A^{\bbz_2} \cup A^{\tilde{\bbz}_2}/2$. 
The gauge invariance of the partition function under $A^{\bbz_2} \to A^{\bbz_2} + d\lambda^{\bbz_2}$ then necessarily implies the existence of an edge mode of the 1-D system.  

If the total symmetry is $G = \bbz'_2 \times U(1)$, the $\bbz_2$ subgroup of $U(1)$ can be gapped so that the symmetry acting on the gapless degrees of freedom is $\bbz'_2 \times \tilde{U}(1)$. If the low energy theory has a mixed anomaly  
\be 
\omega = \frac{1}{2} A^{\bbz'_2}   \cup \frac{d A ^{\tilde{U}(1)}}{2\pi},
\label{eq_alow}\ee
then the anomaly vanishing equation yields
\be 
\alpha  = \frac{1}{2} A ^{\bbz'_2}  \cup A^{\bbz_2}.
\label{eq_alowg}
\ee 
Consequently, the gauge invariance of the partition function requires the existence of an edge mode. The form of $\omega$ and $\alpha$ is very similar to those we discussed in the main text. We also note that the similarity between Eq.(\ref{eq_omeg}) and Eq.(\ref{eq_alow}). 

In the main text, the $\bbz_2$ subgroup of $U(1)$ is the parity $P= \prod_i (-1)^{n_i}$. Unlike the gapping mechanism in Ref.~\cite{Thorngren2021}, in Sec.~\ref{subsec_1d}, $P$ is gapped out (when PBC is used) due to the emergent gauge constraints arising from the terms  $-K \sum \sigma^x_{i-1/2} (-1)^{n_i} \sigma^x_{i+1/2}$ in the Hamiltonian when $K$ is large.  $W = \prod_i \sigma_{i+1/2}^z$, a UV symmetry,  effectively becomes the dual symmetry of $P$. Consequently, there is an emergent mixed anomaly $\omega$ between $W$ and $\tilde{U}(1)$ as in Eq.(\ref{eq_omeg}). On the other hand, we argued in the main text that there is a term $ 
\alpha  =  A^W  \cup A^P/2
$ that dictates the SPT edge modes. $P$ and $W$ being UV symmetries, the anomaly in $\alpha$ when $A^{\tilde{U}(1)}$ is not flat requires it to be canceled by other terms. Indeed, the emergent anomaly $\omega$ serves the purpose if $d A^P = da^{\bbz_2} = \frac{d A^{\tilde{U}(1)}}{2 \pi}\ \mod 2$ where $a^{\bbz_2}$ is the emergent $\bbz_2$ gauge field. Not surprisingly, the anomaly is identical to that in Eq.(\ref{eq_alow}) after we identify $W$ with $\bbz'_2$. This explains the resemblance between Eq.(\ref{eq_omeg}) and Eq.(\ref{eq_alow}).

There is a subtlety if $M$ has boundaries. Boundary conditions need to be chosen properly to guarantee the emergent (dynamical) gauge-invariance. In principle, the true symmetry acting  on the gapless modes are $W$ and $\tilde{U}(1)$, regardless of whether OBC or PBC is used. 
In the main text,  $P$ is interpreted as a UV symmetry, and thus physical. When PBC is used, it is fully gapped. When OBC is used, it also acts nontrivially on the low energy modes. The gauge-invarinace of $ \alpha  =  A^W  \cup A^P/2$ under  $A^P\to A^P + d\lambda^P$ already implies the existence of edge modes when OBC is used, irrespective of the existence of the mixed anomaly. This is because $W$ can terminate on the edges such that $d A^W \neq 0 \mod 2$. It is the delicate cooperation of both $\alpha$ and $\omega$ through 't Hooft anomalies that determines the nature of the intrinsically gapless SPT phase protected by $W$ and $U(1)$ as discussed in Sec.~\ref{sec1d}.

The discussion above about the mixed anomaly between the quotient symmetry and the dual symmetry after gauging a finite subgroup can be generalized to arbitrary dimensions \cite{tachikawa2020}. In $d$-D, the dual $\hat{\bbz}_2$ symmetry is $(d-1)$-form, and the anomaly action is given by $ 
\omega  = \frac{i}{2} \int_Y  A^{\hat{\bbz}_2} \cup   d A^{\tilde{U}(1)}  
$. Here, $A^{\hat{\bbz}_2}$ is a $d$-form background field. If the symmetry $\hat{\bbz}_2$ again coincides with a UV symmetry $W$ and the gauge theory is emergent as we have discussed in the main text, we then again have an SPT phase with 
$ 
\alpha  =  A^{\hat{\bbz}_2}  \cup A^{\bbz_2}/2,  
$ which is again canceled by $\omega$. If both $U(1)$ and $\hat{\bbz}_2$ are preserved, then we may have a higher dimensional intrinsically gapless SPT phase. On the other hand, if $\tilde{U}(1)$ is spontaneously broken as in the Higgs phase, the SPT phase coexists with gapless Goldstone modes.

\section{Perturbations by $\sigma^x$ and $\sigma^z$ in the 1-D Bose-Hubbard model}
\label{app-perturb}
Here we provide further numerical details on the effects of perturbations in the 1-D Bose-Hubbard model. Specifically, we consider two types of perturbations: $h_x\sum_i\sigma^x_{i+1/2}$, which commutes with the effective gauge constraint from the $K$-term but explicitly breaks the $W$ symmetry, and $h_z\sum_i\sigma^z_{i+1/2}$, which preserves the $W$ symmetry but violates the effective gauge constraint.

\begin{figure}[tb]
    \centering
    \includegraphics[width=0.8\columnwidth]{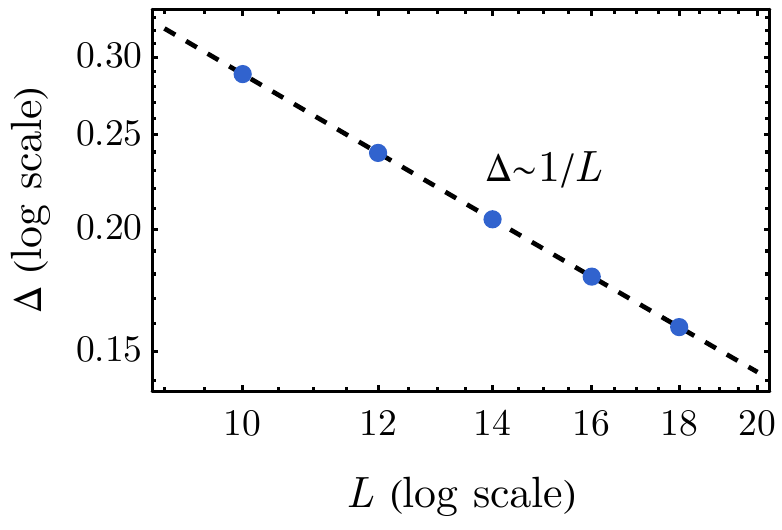}
    \caption{Gap scaling with system size when the $\sigma^x$ perturbation is added. The $1/L$ behavior already shows up for relatively small system sizes. Parameters used: $t=0.5,U=1,h_x=0.1,K=10$.}
    \label{fig:perturb-x}
\end{figure}

\begin{figure}[b]
    \centering
    \includegraphics[width=0.8\columnwidth]{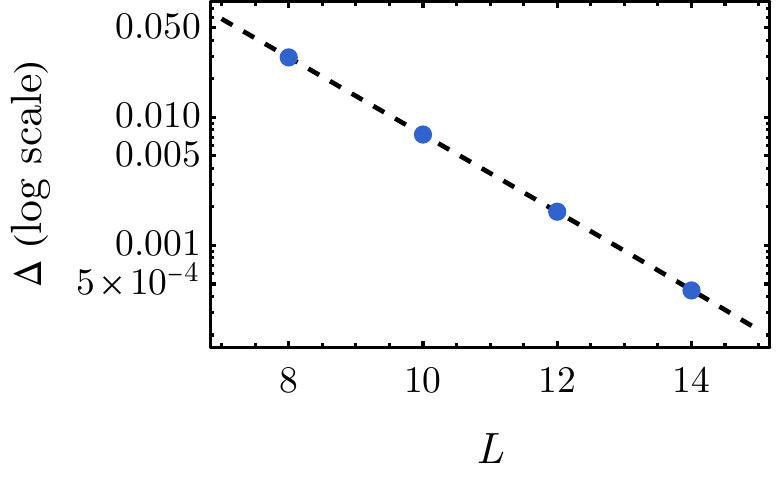}
    \caption{Gap scaling with system size when the $\sigma^z$ perturbation is added. It shows an exponential decay. Parameters used: $t=0.5,U=1,h_z=5,K=10$. Here we called $h_z$ perturbation, but in fact it has to be of the same order with $K$ for a sizable gap to show up even for small system sizes.}
    \label{fig:perturb-z}
\end{figure}

 \subsection{Perturbation by $\sigma^x$}
The exact edge degeneracy in the 1-D case discussed in the main text relies crucially on the existence of the $W$ symmetry. Therefore, we would expect the edge degeneracy to disappear once the $W$-breaking $\sigma^x$ terms are added to the Hamiltonian. Fig.~\ref{fig:perturb-x} shows that by adding even a small such perturbation ($h_x$ much smaller than the other couplings in the Hamiltonian), the degeneracy is immediately lifted, and the gap follows a $1/L$ scaling, with $L$ being the system size. This means the gapless system no longer have edge modes, hence becomes topologically trivial.

 \subsection{Perturbation by $\sigma^z$}
Another interesting type of perturbation is the $\sigma^z$ term. While it preserves the $W$ symmetry, it locally anti-commutes with the effective gauge constraint from the large $K$-term. Recall that the localized action of boson parity on the edges, i.e.,  $P=\sigma^x_{1/2}\sigma^x_{L+1/2}$, derives from the effective gauge constraint being exactly implemented when $K$ flows to infinity at low energy. Therefore, it is reasonable to expect that if there is a sufficiently large gauge-violating term then $P$ will no longer be \textit{exactly} localized on the edges, meaning that $P$ can no longer be separated as two independent $\sigma^x$ operators in a clean manner. Instead, there will be longer range terms effectively that lead to the hybridization of the two edge modes, splitting the exact degeneracy. This intuition is verified by Fig.~\ref{fig:perturb-z}, which further shows that the gap from degeneracy splitting decays exponentially with system size. Therefore, the effective gauge-violating perturbation could lift the exact degeneracy in the fixed point gapless SPT phase and the original exactly localized edge modes now become exponentially localized.

\bibliography{earef}

\end{document}